%% file: sample631.tex
\documentclass[twocolumn]{aastex631}

\usepackage{xcolor}   

\usepackage[T1]{fontenc}
\usepackage{xcolor}
\usepackage{textcomp}

\usepackage{amssymb}
\usepackage{amsmath}
\usepackage{ulem}
\usepackage{physics}
\usepackage{upgreek}
\usepackage{xspace}
\usepackage{comment}
\usepackage{multirow}
\usepackage{array}
\usepackage{booktabs}
\usepackage{etoolbox}

\begin{document}

\newcommand{\grslhaaso}{GRS 1915+105--LHAASO\xspace}
\newcommand{\grs}{GRS 1915+105\xspace}
\newcommand{\grsobs}{LHAASO J1914+1050\xspace}

\def\tsc#1{\csdef{#1}{\textsc{\lowercase{#1}}\xspace}}
\tsc{WGM}
\tsc{QE}
\tsc{EP}
\tsc{PMS}
\tsc{BEC}
\tsc{DE}
\newcommand{\fermi}{\textit{Fermi}\xspace}
\newcommand{\gr}{\ensuremath{\gamma}-ray\xspace}
\newcommand{\grn}{\ensuremath{\gamma} rays\xspace}


\title{Extreme PeV accelerator associated with GRS 1915+105}

\input{ApJ20260713.tex}

\begin{abstract}

Microquasars, binary systems featuring relativistic jets, have emerged as sources for particle acceleration beyond PeV energies. We present a study of the broadband \gr emission from one of the most prominent Galactic microquasars GRS~1915+105 based on data accumulated by LHAASO and \fermi-LAT over 4 and 17 years, respectively. A joint analysis of LHAASO-WCDA and LHAASO-KM2A data reveals extended \gr emission whose centroid appears significantly shifted, by \(\sim 0.13^\circ\), from the binary system and its jets. The spectral energy distribution is well described by a curved spectrum with progressive steepening that can be described by a log-parabola function with no evidence for a sharp cutoff, consistent with parent particles reaching multi-PeV energies and an extreme acceleration efficiency approaching the limit set by the available potential drop across the source. Several features, most notably the shift of the emission and single-power-law spectrum down to GeV band, favor radiation by cosmic rays accelerated in the source interacting with the dense ambient medium. Our spectral modeling implies that at least a few percent of the jet mechanical power is transferred to protons, whose maximum energy reaches beyond  $5$~PeV. These results strengthen the case for microquasars as exceptionally efficient accelerators in our Galaxy.

\end{abstract}




\section{Introduction}

A microquasar is a system in which a compact object (CO) -- such as a black hole (BH) or a neutron star (NS) -- accretes material from a companion star. Unlike typical X-ray binaries (XRBs), microquasars are distinguished by the presence of relativistic jets, which emit non-thermal radiation~\citep{Mirabel:1999fy}. These powerful outflows, whose kinetic power may exceed \(10^{39}\mathrm{\,erg\,s^{-1}}\), are believed to be launched by the extraction of rotational energy from the black hole or by magnetic centrifugal forces arising from the inner accretion disk~\citep{blandford1977electromagnetic, Blandford1982, Meier:2000wk, Fender:2025izt}. Jets of a microquasar are capable of forming extended lobes, the dynamic timescale of which may significantly exceed the duration of the radio/X-ray -astronomy eras, so the absence of observational signatures of a jet in historical data does not rule out the microquasar scenario for a extended structures around an XRB \citep[see][and references therein]{2014SSRv..183..323F}. 

Among stellar-mass black holes, \grs stands out as one of the most dynamic and well-studied microquasars in our Galaxy. First identified as an X-ray source by the WATCH instrument on board the GRANAT observatory in 1992~\citep{castro1992grs}, it quickly became the focus of extensive multiwavelength studies. Subsequent radio observations with the Very Large Array (VLA)~\citep{mirabel1994superluminal, Rodriguez:1998rf}  and MERLIN~\citep{Fender:1998nq} revealed a highly variable radio counterpart with apparent superluminal two-sided ejections. This was the first confirmed instance of relativistic motion detected in a Galactic source, indicating jet speeds approaching the speed of light. Detailed kinematic studies estimated the jet velocity at $v \approx 0.8c$ with an inclination angle of $\theta \approx 63^\circ$ relative to the observer's line of sight~\citep{Rodriguez:1998rf, Fender:1998nq,1999MNRAS.302..253A,2024JHEAp..43...93K}. Observation of the propagation of radio blobs reveals a change of the position angle for different ejection events \citep[for example][]{reid2014parallax}. This can be an indication of the jet precession with an angle of approximately \(\sim10^\circ\).
Recent radio interferometric observations detected even stronger, by an angle of \(\sim35^\circ\), variation in the jet orientation and a misalignment between the continuous jet and discrete ejecta~\citep{Jiang:2026yry, Yan:2026rjj}.
This may indicate a time-variable jet-launching geometry, possibly related to disk warping or strong precession of the inner disk.

The physical properties of \grs have been refined through extensive observations. 
Initial distance estimates provided by~\cite{reid2014parallax}, were later improved with 3D kinematic measurements, yielding a refined distance of $d = 9.4 \pm 0.6 \pm 0.8 \,\text{kpc}$~\citep{Reid:2023ksq}. 
The black hole mass, initially proposed to be $M_\mathrm{BH} \approx 14 M_{\odot}$~\citep{Greiner:2001ie}, was later revised to values ranging between $10M_{\odot}$ \citep{Steeghs:2013ksa} and $12M_{\odot}$~\citep{reid2014parallax, Reid:2023ksq}. The corresponding Eddington luminosity is estimated to be on the order of $L_\mathrm{Edd} \simeq 1.3 \times 10^{39} (M_\mathrm{BH} / 10 M_\odot) \, \text{erg} \, \text{s}^{-1}$. Modeling of the radio-bright ejection from \grs revealed that the jet power should be close or even exceeding this limit \citep{atoyan1999modelling, 2024JHEAp..43...93K}, placing it among the most powerful X-ray binaries known.

The donor star in this system has been identified as a K-type star \citep{Greiner:2001ie}, with a mass of approximately $0.5 M_{\odot}$ \citep{Steeghs:2013ksa}, confirming \grs as a low-mass X-ray binary (LMXB). Additionally, its orbital period of $P_{\rm orb} = 33.85 \pm 0.16$ days \citep{Steeghs:2013ksa} provides critical constraints on the system's evolution.

Long-term multiwavelength observations have revealed significant variability in \grs across different energy bands. The system exhibits quasi-periodic oscillations (QPOs) in both the radio ~\citep{Pooley:1997my, Fender:2000fa} and the infrared bands~\citep{Fender:1998bb}, indicating dynamic interactions between the accretion disk and the jet. In addition, \grs frequently undergoes episodic flaring, with notable anti-correlations observed between X-ray and radio emissions, suggesting an alternating energy distribution between the inner accretion flow and the relativistic jets~\citep{Mendez:2022wpu}.

For more than two decades, \grs remained a persistently bright X-ray source, maintaining luminosities between $0.1 \text{--} 1 \, L_{\rm Edd}$. However, in June 2018, the system underwent an unexpected and dramatic dimming, with the X-ray flux dropping by over an order of magnitude. This decline was accompanied by increased flaring activity at longer wavelengths, suggesting a fundamental shift in the accretion state of the system. The physical origin of this transition remains a subject of debate, with scenarios ranging from heavy obscuration by a local outflow~\citep{Motta:2021htt, Balakrishnan:2020yuo} to an intrinsic transition to a low-luminosity hard state~\citep{Motta:2021htt, Koljonen:2021dqn}.

High-energy (HE) and very-high-energy (VHE) \gr emissions has long been predicted from microquasar jets~\citep{aharonian2004very,2005A&A...432..609B, Bosch-Ramon:2008xrf}.
With \grs being one of the most extreme Galactic microquasars, which jet power exceeds \(10^{39}\mathrm{erg\,s^{-1}}\) \citep{atoyan1999modelling,Motta:2025lgb}, prior to LHAASO no \gr signal had been detected from this system~\citep{abdallah2018search}. A tentative signal reported by HEGRA~\citep{Aharonian:1996jr} could not be confirmed with H.E.S.S.~\citep{abdallah2018search} or MAGIC~\citep{MAGIC:2009rke}, which can be interpreted either as a manifestation of the variability or as the low significance of the source detection with HEGRA. LHAASO collaboration reported on a highly significant, \(\approx 15.1\sigma\), detection of \gr emission from the source with KM2A  (above \(25\)~TeV, i.e., confirming it as an ultra-high-energy --UHE-- source \citep{LHAASO:2024psv}). Following this detection, analysis of \fermi-LAT data revealed a persistent multi-GeV emission consistent with the position of \grs \citep{2025ApJ...979L..40M}.  

In this work, we extend the analysis of the LHAASO data in \cite{LHAASO:2024psv} to lower energies by incorporating the Water Cherenkov Detector Array (WCDA). Furthermore, longer exposure of Kilometer Square Array (KM2A) allowed us to confirm detection up to \(1\)~PeV energies. 
We also revisit the \fermi-LAT data focusing on the region bright in the VHE and UHE bands.

\section{LHAASO data and analysis}\label{sec:data}
In this analysis, we use full-array data from the WCDA collected between 5 March 2021 and 31 July 2025, and from the KM2A collected between 20 July 2021 and 31 July 2025. The corresponding effective live times are 1484.4 and 1438.8 days, respectively. To ensure reliable event reconstruction, we restrict the analysis to events with zenith angles smaller than $50^\circ$.

For WCDA, events are grouped into six bins based on the energy estimator defined by the number of triggered detector cells, $N_{\rm hit}$, with ranges of [60--100), [100--200), [200--300), [300--500), [500--800), and [800--2000]. These bins correspond approximately to true energies from about 300~GeV to 30~TeV, although the exact mapping depends on the assumed spectral shape and the source declination. For KM2A, the reconstructed energy $E_{\rm rec}$ is used as the energy estimator, and events are grouped into ten logarithmic bins with a width of $\Delta \log_{10} E_{\rm rec} = 0.2$, covering energies from 10~TeV up to 10~PeV.
The sky map is constructed in the equatorial coordinate system with a grid resolution of $0.1^\circ \times 0.1^\circ$, where each pixel records the number of detected events based on their reconstructed arrival directions. The cosmic-ray background in each pixel is estimated using the direct integration method~\citep{Fleysher:2003nh}, after masking the Galactic plane and all detected sources. Prior to filling the sky map, events are subjected to gamma-hadron separation cuts to suppress the majority of the cosmic-ray background.

The region of interest (ROI) is rectangular centered the position of \grs, covering right ascension from $286.6^\circ$ to $290.6^\circ$ and declination from $9^\circ$ to $13^\circ$. Resolving sources is performed using an iterative binned Poisson maximum-likelihood method. The test statistic is defined as $TS = -2 \ln (\mathcal{L}_0 / \mathcal{L}_1)$, where $\mathcal{L}_1$ and $\mathcal{L}_0$ are the likelihoods for the signal-plus-background and background-only hypotheses, respectively. Sources are added iteratively until the increase in $TS$ becomes statistically insignificant ($\Delta TS < $ 12--18 for 4--7 additional free parameters).  We adopted here a low TS threshold to include sub-threshold sources in the fit, and this approach adheres to information-theoretic principles by minimizing information loss, ultimately leading to more conservative and robust flux estimates. The expected signal counts in each pixel are calculated by summing the contributions from all sources, obtained through forward-folding the assumed spectral model with the detector response, including the detection efficiency, live-time distribution, source extension model, and point-spread function (PSF). Source positions, spectral parameters, and extensions are treated as free parameters and are determined via maximum-likelihood optimization.

WCDA and KM2A data are analyzed jointly by summing the likelihood contributions over all $N_{\rm hit}$ (WCDA) and $E_{\rm rec}$ (KM2A) bins, assuming that the model parameters are consistent across the full energy range. Prior to the analysis, the relative
 pointing offsets and point-spread functions of both detectors, as well as their relative energy scale, are verified, calibrated, and corrected using well-established point sources in neighboring declination bands to ensure the reliability of the joint analysis of WCDA and KM2A.

In the fitting procedure, several spectral models, including a power law (PL), a log-parabola (LP), and a power law with an exponential cutoff (PLEC), are tested iteratively for each individual source. The best-fitting model is selected using the Akaike Information Criterion (AIC)~\citep{AIC}. 

The sources may exhibit extended emission. Three scenarios are considered for estimating the source extension: a point-source model, a Gaussian template, and an energy-dependent Gaussian template in which the extension $\sigma_\textrm{ext}$ varies with energy $E$ as $\sigma_\textrm{ext} = \sigma_\textrm{ext,0} + b \log_{10}(E/E_{0})$, where $E_0 = 10\,\textrm{TeV}$ is the pivot energy, $\sigma_\textrm{ext,0}$ denotes the extension at $E_0$, and $b$ is the coefficient describing the energy dependence. The linear dependence on $\log_{10}E$ represents the simplest assumption, as more complex parameterizations are generally not supported by the available statistics. The optimal extension scenario is generally selected using the AIC. However, for comparisons involving nested models, a dedicated treatment is adopted. In such cases, we rely on the profile likelihood~\citep{Rolke2005_NIMA551_493}, which provides a more appropriate assessment of the statistical significance of additional parameters. 

In addition to discrete sources, the Galactic diffuse emission (GDE) within the ROI is non-negligible and plays an important role in the analysis. We adopt the dust column density measured by the \textit{Planck} satellite~\citep{Planck2013DustModel} as the spatial template and model the spectrum with a smooth broken power-law function $\dd N/\dd E = F_0 \left(E/E_0\right)^{-\alpha} \left[1 + \left(E/E_\textrm{b}\right)^{\omega}\right]^{(\alpha-\beta)/\omega}$, where $E_\textrm{b}$ denotes the break energy, the sharpness parameter is fixed to $\omega = 5$, $\alpha$ and $\beta$ are the spectral indices before and after the break.
To accurately determine the flux and spectral parameters of the GDE, the ROI is extended along Galactic latitude to $\pm 5^\circ$ and fitted by iteratively adding point-like and extended sources. In the enlarged ROI, the dust distribution at the latitude boundaries provides important constraints on the GDE SED fitting, thereby reducing the relative influence of individual sources. This approach allows the large-scale, dust-associated GDE to be modeled while minimizing the impact on small-scale local structures. The resulting best-fit flux normalization of the GDE is then scaled to the nominal ROI, and both the normalization and the spectral parameters of the GDE are fixed in the subsequent re-analysis of the nominal ROI.

Details of the analysis are provided in Supplementary Section~\ref{sec:supanalysis}. In total, four sources are identified within the ROI. The detailed properties of these sources are listed in Supplementary Table~\ref{tab:roi_sources}.

\begin{figure}
    \centering
    \includegraphics[width=0.45\textwidth]{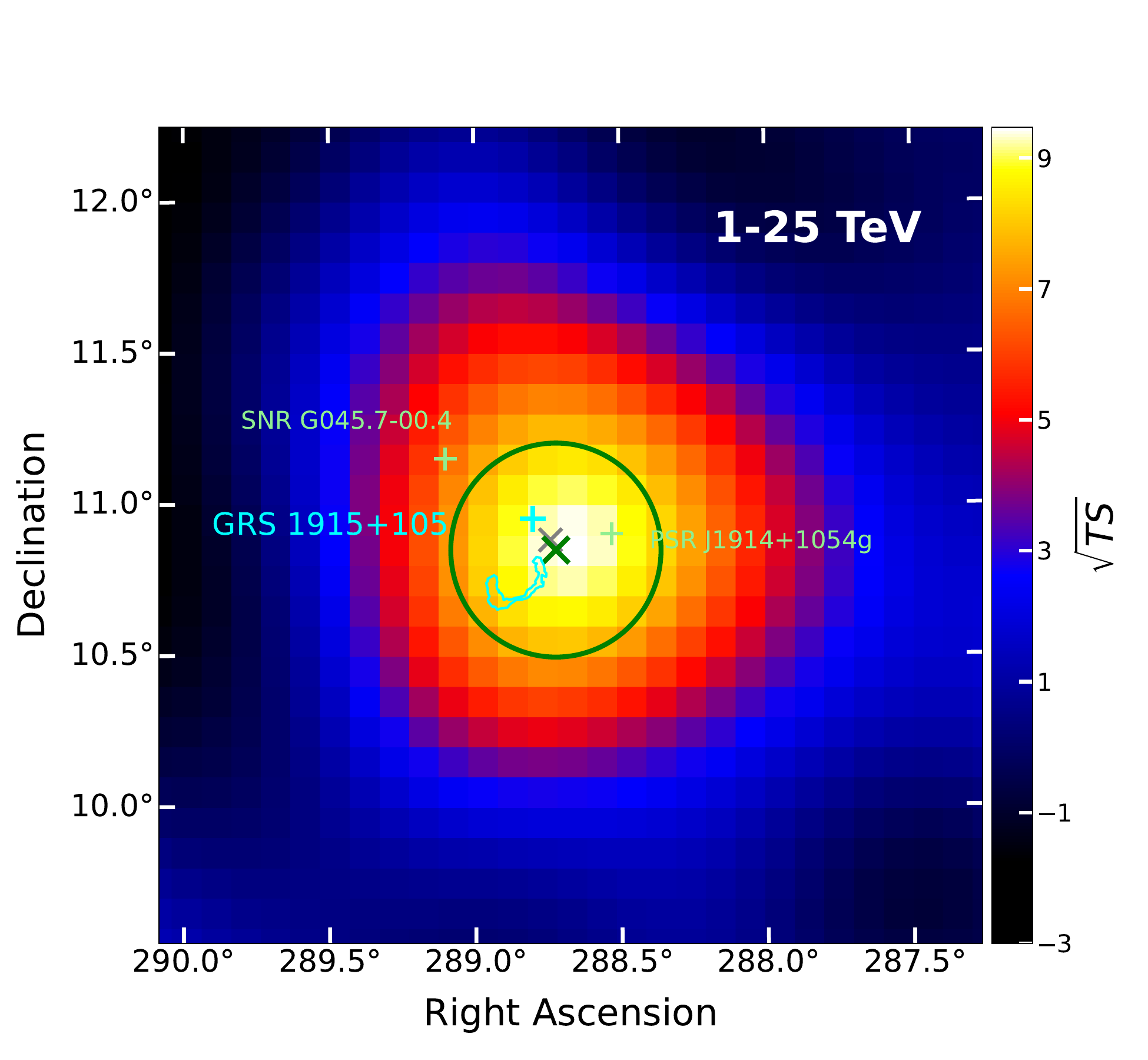}
    \includegraphics[width=0.45\textwidth]{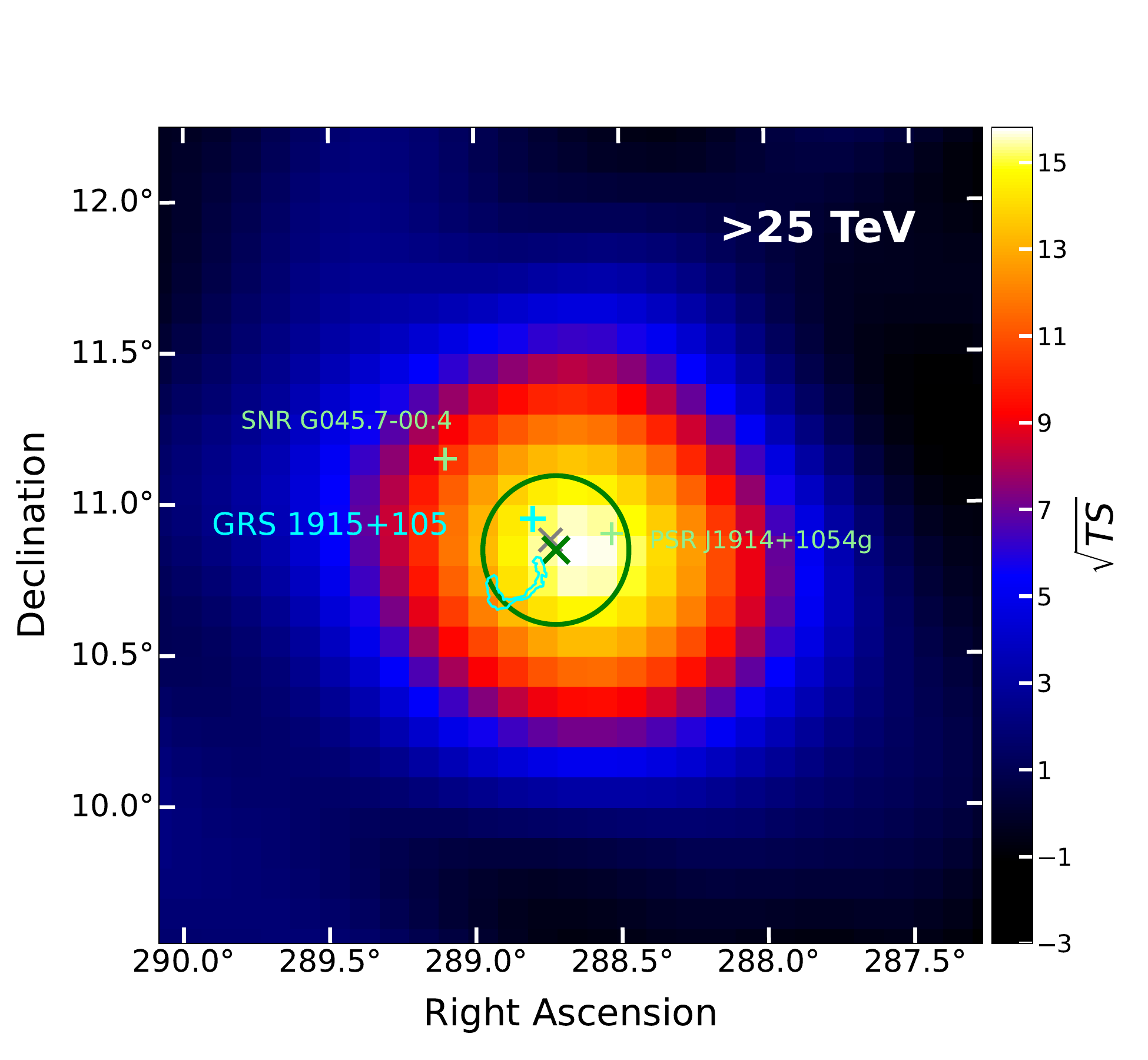}
    \caption{Significance maps of \grsobs (\grslhaaso) observed with LHAASO-WCDA (upper panel) and with LHAASO-KM2A (lower panel). The dark-green circles indicate the measured source extension, centered on the dark-green cross ($\times $). The radii of the circles are extension at $5\,\textrm{TeV}$ (0.354$^{\circ}$) and that at $100\,\textrm{TeV}$ (0.246$^{\circ}$). The cyan plus ($+$) symbol marks the nominal position of \grs, while the gray cross ($\times$) shows the source position reported in the previous LHAASO-KM2A measurement~\citep{LHAASO:2024psv}. Light green plus ($+$) symbols denote the positions of other known supernova remnants (SNRs) or pulsars, and cyan contour marks the jet-interstellar medium (ISM) interaction region observed by MeerKAT~\citep{Motta:2025lgb}.}
    \label{fig:sigmap}
\end{figure}

Among these sources, \grsobs, is detected with a total significance of $17.7\,\sigma$ ($9.5\,\sigma$ and $15.8\,\sigma$ in 1--25\,TeV and > 25\,TeV, respectively, see Fig.~\ref{fig:sigmap}), making it one of the most significant detections in the region. It is located at RA $= \left(288.722^{+0.036}_{-0.036}\right)^{\circ}$ and Dec $= \left(10.846^{+0.037}_{-0.036}\right)^{\circ}$, exhibiting significant extension. 
The energy-dependent Gaussian spatial extension of
\begin{align}\label{eq:lhaaso_ext}
   \sigma_{\rm ext} &= \left(0.329^{+0.049}_{-0.051}\right)^\circ \nonumber\\
   &- \left(0.083^{+0.051}_{-0.055}\right)^\circ\times\log_{10}\left(\frac{E}{10\,\textrm{TeV}}\right). 
\end{align}
is preferred as the $1\,\sigma$ confidence interval of $b$ excludes zero. The emission centroid, which shows no energy-dependent shift in separate WCDA and KM2A analyses, lies at an angular separation of $0.13^\circ$ from the nominal position of \grs, and within $0.05^\circ$ of the position reported in the previous LHAASO study~\citep{LHAASO:2024psv} in which the measured extension is $\sigma_{\rm ext} = 0.28 \pm 0.05^\circ$ (at $E\ge25\,\textrm{TeV}$).
The bow shock structure discovered by MeerKAT is well-aligned with \grsobs~\citep{Motta:2025lgb}.

\begin{figure}
    \centering
    \includegraphics[width=1.0\linewidth]{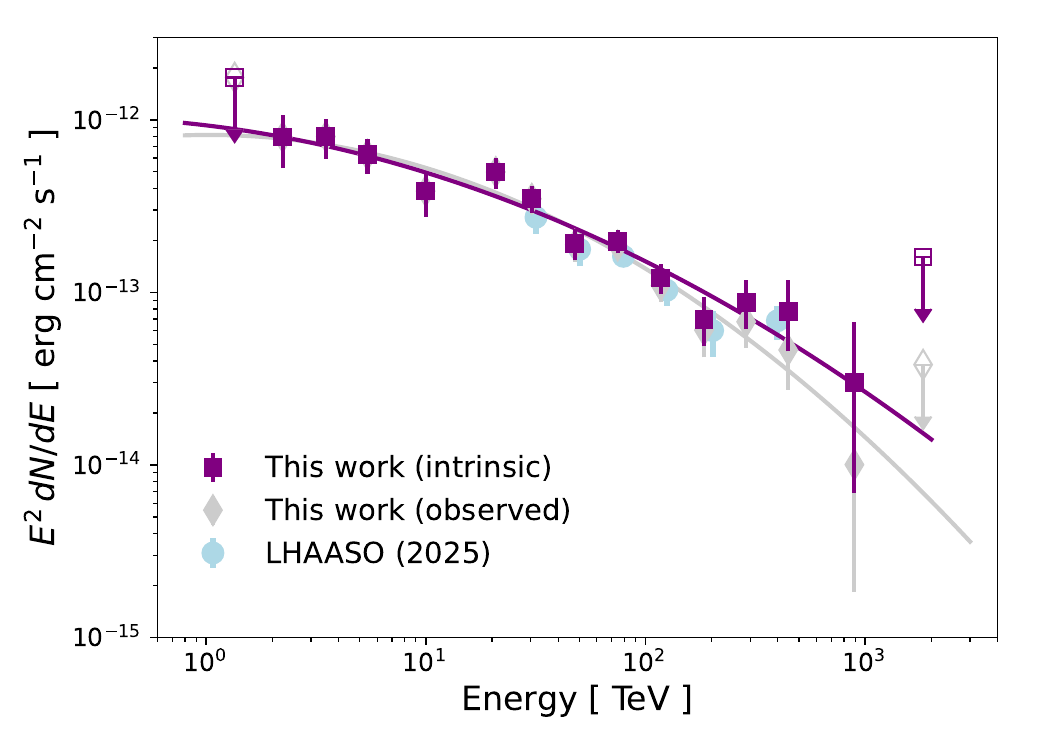}
    \caption{VHE–UHE SED of \grsobs (\grslhaaso) measured by LHAASO, modeled with a LP (log-parabola) function. The purple and light gray data points represent the de-absorbed (corrected for IR and CMB attenuation) and observed SEDs, respectively, while the solid lines show the corresponding best-fit models. The light blue points are published results from \cite{LHAASO:2024psv}.}
    \label{SED}
\end{figure}

The spectral energy distribution (SED) of \grsobs is well described by a log-parabola model, $\dd N / \dd E = F_0 \left(E/E_0\right)^{-\alpha - \beta \log(E/E_0)}$, with the flux normalization $F_0 = \left(3.09^{+0.70}_{-0.65}\right)\times10^{-15}\,\textrm{TeV}^{-1}\textrm{cm}^{-2}\textrm{s}^{-1}$, the spectral index $\alpha = 2.394^{+0.095}_{-0.122}$ at the pivot energy ($E_0 = 10\,\textrm{TeV}$) and the curvature parameter $\beta = 0.052^{+0.036}_{-0.031}$. The best-fit spectrum and corresponding data points are shown in Fig.~\ref{SED}, where models with and without absorption by infrared (IR) and cosmic microwave background (CMB) photons are compared. In both cases, no sharp cutoff is observed up to energies of $\sim$1\,PeV.

\grsobs shows good agreement in position, extension, and spectrum with the results of the previous LHAASO analysis of \grs~\citep{LHAASO:2024psv}.

To further assess possible alternative counterparts, we individually added each known pulsar and supernova remnant, as well as sources listed in the \fermi-LAT Fourth Source Catalog (4FGL) within $0.5^\circ$ of \grs, with their positions fixed during the fitting procedure, and found no significant increase in the summed TS value (see Supplementary Section~\ref{subsec:SNR_PWN_influence}). Thus, \grs remains the most plausible counterpart of the detected emission. Hereafter, unless otherwise stated, \grslhaaso refers to the extended \gr emission detected by LHAASO and labeled as \grsobs in this analysis.

Additional tests of SED models (e.g., the SBPL model) and morphology models (e.g., elliptical Gaussian model) were performed for the target source \grslhaaso. No improvement in TS values was found when evaluated using the AIC (see Supplementary Section~\ref{subsec:sed_mor}). This provides evidence for the stability of the analysis. An analysis of the flux variability was performed, and no hint of flux variation was found over the period studied.

\section{\fermi-LAT observations of \grslhaaso}

\begin{figure}
\centering
      \includegraphics[width=1.0\linewidth]{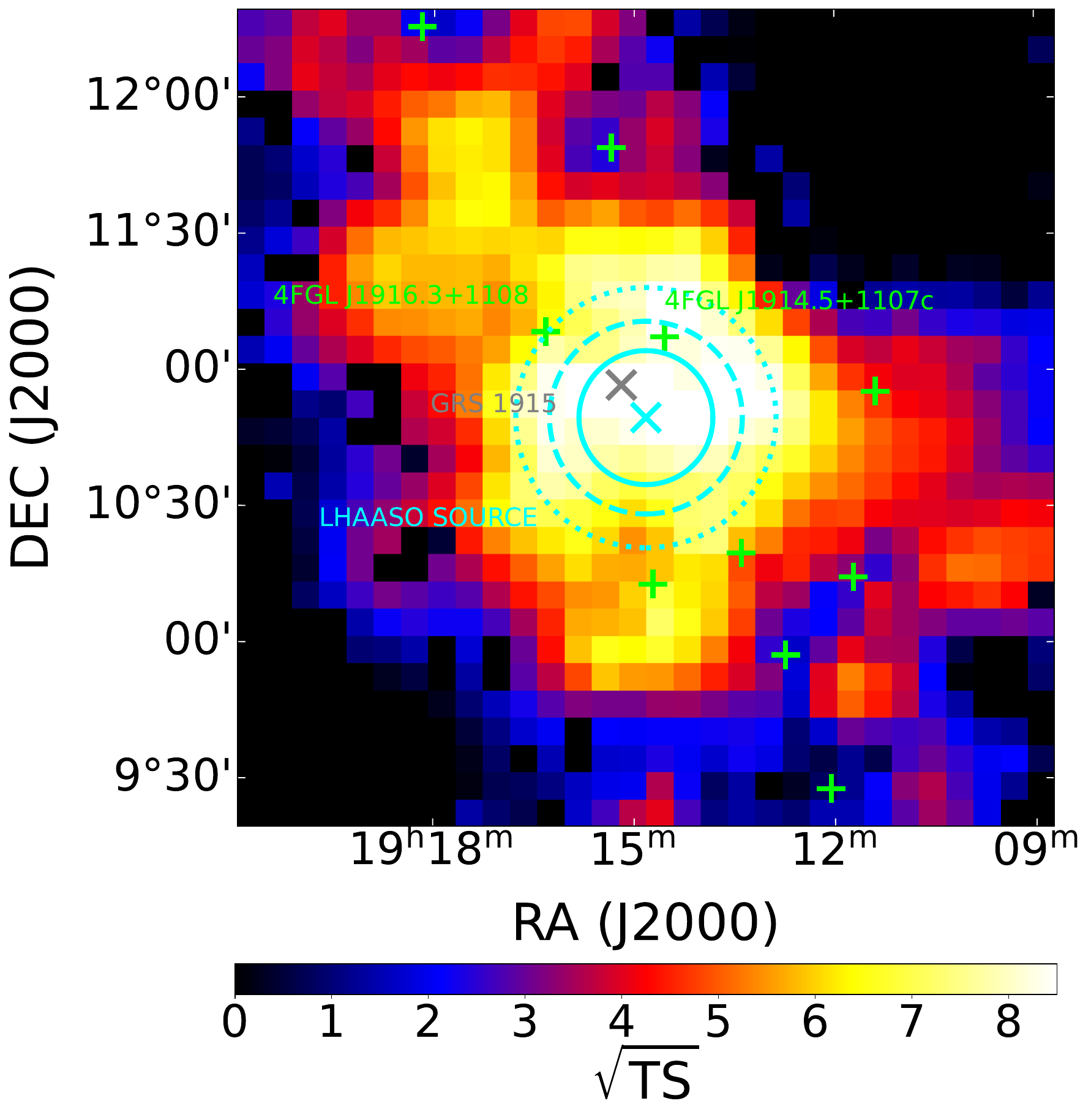}
      \caption{1--500~GeV significance map for the region around \grslhaaso. All catalog sources were considered and removed. The green pluses mark the positions of the catalog sources, the gray cross marks the position of \grs. The cyan cross marks the center position of \grslhaaso, and the cyan solid, dashed and dotted circles represent the extension obtained at 100~TeV, 5~TeV, and from the \fermi\ analysis.}
      \label{fig:tsmap_main}
\end{figure}

With the data obtained from the Large Area Telescope (LAT) onboard the \fermi\ Gamma-ray Space Telescope, we analyzed the \gr\ emission in the sky region of \grslhaaso. Using 0.1--500~GeV \fermi-LAT data accumulated over 17 years, we performed a likelihood analysis (see Supplementary Section~\ref{sec:si}). We detected a significant GeV excess spatially coincident with \grslhaaso. When modeled with the LHAASO-based spatial template, the source was detected with a TS of 223. The \gr\ emission is well described by a power-law spectrum with a photon index $\Gamma = 2.07 \pm 0.04$ and a 0.1--500~GeV photon flux of $(3.4 \pm 0.5) \times 10^{-8} \rm\,photons\,cm^{-2}\,s^{-1}$. We searched for any
long-term variability of the \gr\ emission by calculating the variability 
indice TS$_{var}$ following the procedure 
introduced in \cite{nol+12}. The computed TS$_{var}$ is lower than the threshold for identifying a variable source (see Supplementary Section~\ref{sec:vi}), indicating no significant long-term variability. The 1--500~GeV significance map is shown in Fig.~\ref{fig:tsmap_main}.

Since \grslhaaso\ lies in a crowded region and a prior source template was adopted, we carefully investigated possible contamination from nearby \gr\ sources, in particular the brightest nearby source 4FGL J1916.3+1108 (see Supplementary Section~\ref{sec:contamination}). This source is listed as a point source in the \fermi\ source catalog, but has been suggested to be possibly extended and associated with the supernova remnant G045.7--00.4~\citep{Zhang:2021jeq,2025ApJ...979L..40M}. Modeling this nearby source as an extended source improved the global fit, but did not eliminate the need for a \gr\ emission component associated with \grslhaaso. 
In this case, we obtained $\Gamma = 2.00 \pm 0.05$ and a flux of $(1.9\pm 0.3) \times 10^{-8}\rm\, photons\, cm^{-2}\, s^{-1}$ for \grslhaaso, with a TS of 103.

We checked the spatial morphology of the \gr\ emission and confirmed that it favors an extended morphology. A Gaussian disk with a best-fit radius of $\left(0.478^{+0.043}_{-0.038}\right)^{\circ}$ was obtained from the likelihood analysis (see Supplementary Section~\ref{sec:sda}). With this best-fit morphology, we obtained a spectral index $\Gamma = 2.03 \pm 0.04$ and a 0.1--500~GeV photon flux of $(4.4 \pm 0.6) \times 10^{-8}\rm\, photons\, cm^{-2}\, s^{-1}$ for \grslhaaso, with a TS value of 261. 
The \gr\ spectrum obtained with the best-fit model is plotted in Fig.~\ref{fig:GRS1915-SED-fit}. We also checked the possible emission contributions from the other three LHAASO-detected sources (see Supplementary Table~\ref{tab:roi_sources}) and found that the best-fit extension is robust and shows no significant change (see Supplementary Section~\ref{sec:contribution}). The \gr\ spectra derived with different source models are also consistent with each other within uncertainties.

\section{Modeling and interpretation}\label{sec:model}

In this section, we explore the physical origin of the broadband \gr emission from \grs, {using the spectral properties and morphology revealed with LHAASO. As it was shown for the case of SS~433, \(\gamma\)-ray morphology provides a critical tool to discriminate between leptonic and hadronic scenarios. The energy-depended morphology revealed with H.E.S.S. below \(30\)~TeV strongly favors IC emission by relativistic electrons~\citep{HESS:2024rlh}. In contrast, the UHE emission detected with LHAASO in the region off the jet axis, requires hadronic component~\citep{LHAASO:2024psv}. For an extended Galactic source, proton-proton interactions ($pp$) are the dominant hadronic mechanism for production of UHE emission. In this scenario, UHE protons inelastically interact with target protons from the ambient gas. Neutral pions (\(\pi^0\)) created at such collisions decay into \grn.

The morphology revealed with LHAASO in the range from a few TeV to PeV energies appears shifted from the jet axis and the bow shock structure~\citep{Motta:2025lgb}. We therefore  focus on the hadronic scenario in the main text, keeping the discussion of the IC leptonic scenario and related challenges to Supplementary Section~\ref{app:leptonic}. } 
We approximate the distributions of relativistic protons by a standard power-law distribution with an exponential cutoff (PLEC) model: 
\begin{equation}
\dd N/\dd E = A E^{-s_p} \exp(-E/E_{p,\rm max}).
\end{equation}
The numerical value of spectrum normalization constant, \(A\), is non-intuitive, 
for the sake of physical reasoning we present the normalization parameter \(A\) as the total energy stored in the distribution, $W_p \equiv \int_{E_{p,\rm min}}^{\infty} E \dd N$.

We perform SED fitting using LHAASO-only (orange line) and combined LHAASO+LAT (blue line)  datasets for $pp$ scenario, the results of which are presented in Fig.~\ref{fig:GRS1915-SED-fit}.
The corresponding posterior distributions are shown in Supplementary Figs.~\ref{fig:corner-plot-IC-pp}--\ref{fig:corner-plot-IC-pp-LHAASO}, and the median values, together with their $1\sigma$ uncertainties and the adopted prior ranges, are listed in Supplementary Tables~\ref{tab:fit_params_fermi}--\ref{tab:fit_params}. 

Our spectral fitting for the LHAASO-only dataset  
yields a total proton energy of 
\begin{equation}
W_p \approx 14.9_{-9.1}^{+21.8} \times 10^{49}\, \qty(n_{\mathrm{gas}}/100~\mathrm{cm^{-3}})^{-1}\,\mathrm{erg}, 
\end{equation}
with parameters $s_p \approx 2.5\pm0.1$ and $E_{p,\mathrm{max}} \approx 10.1_{-6.9}^{+32.3}\,\mathrm{PeV}$.
When the fit is extrapolated to the GeV band, the flux uncertainty is large, although marginally consistent with the Fermi-LAT flux level. 
When Fermi-LAT data are included in the fit, the inferred parameters become 
\begin{equation}
W_p \approx 7.9_{-2.4}^{+2.7} \times 10^{49}\qty(n_{\mathrm{gas}}/100\,\mathrm{cm^{-3}})^{-1}\rm~erg, 
\end{equation}
with a spectral index $s_p \approx 2.4\pm0.05$ and a cutoff energy of $E_{p,\textrm{max}} \approx 5.1_{-2.8}^{+12.7}\rm~PeV$. 
We note that the estimate for \(W_p\) depends on the minimal energy of non-thermal protons as \(\propto E_{p,\rm min}^{-1/2}\). Protons with energy below \(10\)~TeV have a minor impact on the spectrum in the LHAASO band, thus the required energy budget can be reduced by a factor of \(\approx10^2\) adopting \(E_{p,\rm min}\approx 10\)~TeV. 

\begin{figure}
    \centering
    \includegraphics[width=1.0\linewidth]{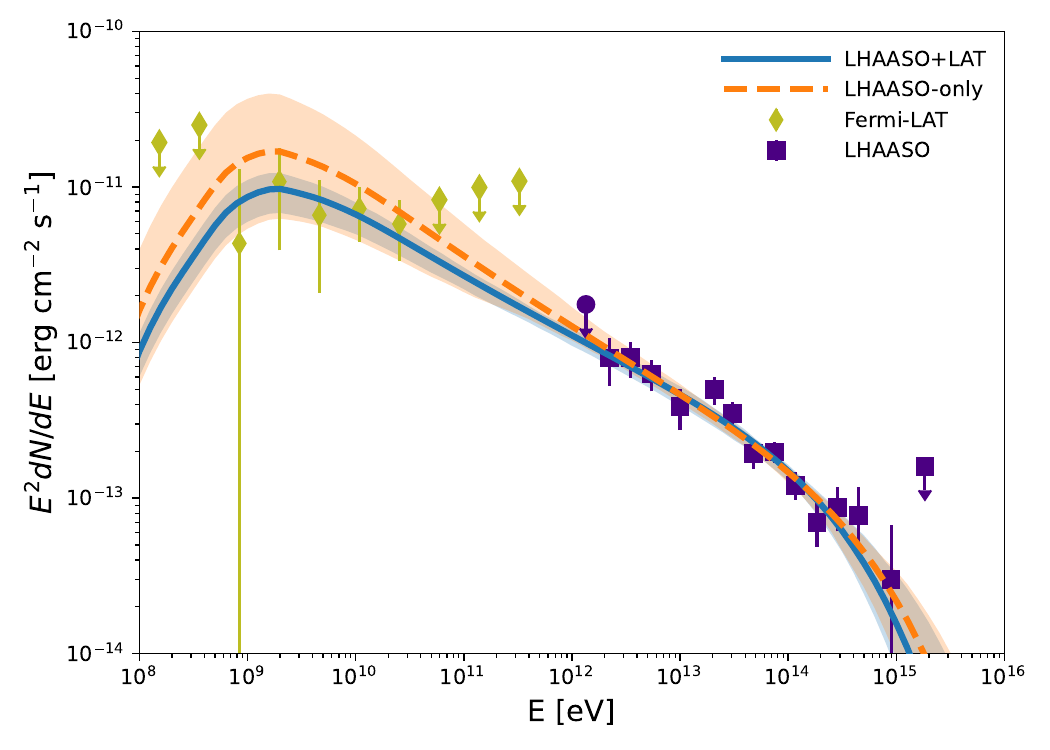}
    \caption{Spectral energy distributions of \grs under 
    hadronic ($pp$) scenarios. The dashed curves show the results obtained using LHAASO data only. 
    }
    \label{fig:GRS1915-SED-fit}
\end{figure}

The energy loss timescale for protons via the $pp$ process is 
\begin{equation}
t_{pp} \approx (\sigma_{pp} \kappa_{pp} n_{\rm gas} c)^{-1} \simeq 0.5 \, (n_{\rm gas}/100\,\text{cm}^{-3})^{-1}\,\text{Myr},
\end{equation}
assuming a cross-section $\sigma_{pp} \simeq 4.5\times 10^{-26}\,\text{cm}^2$ and an inelasticity coefficient $\kappa_{pp} \simeq 0.5$. Note that this cross-section is 50\% larger than the nominal value for GeV protons. The age of the jet is estimated to be $t_{\text{age}} \approx 0.09$--$0.2\,\text{Myr}$, with an upper limit of $0.4\,\text{Myr}$~\citep{Motta:2025lgb}. Since $t_{pp} > t_{\text{age}}$, the required proton injection rate can be estimated as $Q_p \approx W_p / t_{\text{age}} \simeq 2.5\times 10^{37}\,(n_{\rm gas} / 100\,\text{cm}^{-3})^{-1}\,\text{erg s}^{-1}$.

In ISM the ambient gas density may vary significantly. The averaged gas density close to the Galactic disk is about $1\,\mathrm{cm^{-3}}$ and in the center of dense molecular clouds it may achieve \(10^4\,\mathrm{cm^{-3}}\). This range critically influence the power required to support the emission. If the region IRAS 19132+1035, located apparently at the apex of the jet -- ISM interaction, can be treated as a representative probe for the entire complex, then the ISM density in the LHAASO source is very high, \(n_{\mathrm{gas}}\approx 160\,\rm~cm^{-3}\) \citep{Motta:2025lgb}. On the other hand, the averaged density in the LHAASO source apparently cannot be much lower than this value, \(n_{\mathrm{gas}}\gtrsim10\rm~cm^{-3}\) given the constraints on the emission obtained from the large-scale structure \citep{Motta:2025lgb}. Therefore, the obtained above value of \(\sim10^{37}\,\mathrm{erg\,s^{-1}}\) seems to reflect the required accelerator power. 
The available observational data constrain  mechanical power of the jet, \(L_{\rm jet}\), in the range 
$L_{\rm jet} \sim 3.3 \times 10^{37} \text{--} 1.5 \times 10^{39}\,\rm~erg~s^{-1}$, with uncertainties raising mainly from the unconstrained opening angle of the jet~\citep{Motta:2025lgb}. Thus, if a fraction larger than a few percent of the jet kinetic power is transferred to cosmic rays, then the accelerator is \grs can support the \(\gamma\)-ray emitting population of protons.

Below, we estimate the maximum energy achievable by particles accelerated by the jet in \grs.
The acceleration timescale is parametrized as $t_{\rm acc} =  r_L/(\eta_{\rm acc}c)$ with $r_L = E/ZeB$ is the Larmor radius, $\eta_{\rm acc} \lesssim 1$ represents acceleration efficiency. 
The diffusion coefficient in the acceleration site is parametrized as $D = \chi D_{\rm bohm}\approx \chi E_{e} c /(3eB)$.
The size of the acceleration region is constrained by the confinement time $R > \sqrt{6 D t_{\rm acc}}$.
The maximum energy of CR nuclei is 
\begin{equation}
E_{\rm max} \simeq 5.2 Z \eta_{\rm acc,-1}^{1/2} \chi^{-1/2} \sigma_{\rm jet, -1}^{1/2} L_{{\rm jet}, 39}^{1/2} \beta_{-0.1}^{-1/2}\rm~PeV,
\end{equation}
where $Z$ is nuclear charge number, $\sigma_{\rm jet}$ is the magnetization parameter. 
The bulk jet speed is expected to be mildly relativistic with estimates ranging from  $\beta \gtrsim 0.9$ (in units of light speed)  for fast transient jet \citep{Fender:2025izt} down to approximately \(\beta\approx0.5\) on the parsec scale \citep[see][and references therein]{2024JHEAp..43...93K}. 
Thus, particle acceleration to multi-PeV is possible if one assumes a very efficient acceleration process, \(\eta_{\rm acc}\gtrsim10^{-2}\).

Detailed analysis of the TeV \gr data reveals a marginal (but potentially very important) effect of the source size shrink with increase of \gr energy. Such behavior is typically expected in leptonic sources as higher energy electrons lose their energy faster. In case of hadronic scenario, one expects a larger extension of higher energy emission, as the source size is dominated by the particle transport. However, the revealed energy-dependent morphology could also be caused by a recent powerful outburst in the source. Such an outburst would result in the smaller extension of the higher-energy emission as well as in tentative hump seen in the SED in Fig.~\ref{fig:GRS1915-SED-fit} (compare to Fig.~12 in Ref.~\citep{2005A&A...432..609B}).

\section{Discussion and implications}
In this study, we report on the detection of the VHE \gr emission from \grs. Using the joined analysis of WCDA and KM2A data, we obtained the spectrum in a broad energy range from TeV to PeV. 
We also analyzed the \fermi-LAT data from the region inferred with LHAASO observations extending the spectral range down to GeV energies. The apparent shift of the \gr emission from the binary system and the jet termination regions favors the hadronic $pp$ scenario. The spectrum and flux of \grs measured with LHAASO can be explained by protons accelerated in the jet of \grs provided that the acceleration efficiency is high, \(\eta_{\mathrm{acc}}>10^{-2}\), and cosmic rays obtain at least a few percents of the jet kinetic power.

Our spectral modeling suggests that \gr emitting protons follow a power-law energy distribution with the spectral index of \(\approx2.4\). This index is determined by a combined effect of the acceleration and propagation processes. For typical diffusion regimes, such spectrum implies a hard injection of accelerated protons, close to the DSA nominal value of \(2\). Such hard spectral index is required to explain the observed PeV proton spectrum  \citep{LHAASO:2025byy, Aharonian:2026tzf}. 

These findings support the hypothesis that microquasars are Galactic PeVatrons,  i.e., a source population that supply cosmic rays in the PeV energy range~\citep{LHAASO:2024psv}. 
The required power to sustain the density of PeV cosmic rays in the entire Galaxy is around $10^{38}\rm~erg~s^{-1}$.
This means that a few to order of tens similar to \grs microquasars are sufficient  to explain the observed PeV cosmic rays~\citep{LHAASO:2024psv, Zhang:2025tew, Wang:2025yqy}.
Future detailed modeling and/or observations, in particular in the X-ray band, with higher resolution IACTs \citep{2025Resea...8..872C},  or possible detection with next generation neutrino telescopes, will help to develop comprehensive models for this fascinating source. 

\section*{Conflict of interest}
The authors declare no competing interests.

\section*{Acknowledgments}
The LHAASO Observatory (CSTR: \href{https://cstr.cn/31117.02.LHAASO}{31117.02.LHAASO}), including its detector systems, was designed and constructed by the LHAASO project team and is operated and maintained by the LHAASO operations team. We sincerely thank all members of both teams, with special appreciation for those who work year-round at the LHAASO site at an altitude exceeding 4,400 meters. Their sustained dedication ensures the reliable operation of the detector systems and essential infrastructure, including the power supply.

We sincerely acknowledge the Chengdu Management Committee of Tianfu New Area for its sustained financial support of research based on LHAASO data. We also thank the National High Energy Physics Data Center for providing the computing resources and data services that made the analysis in this work possible.

This work was supported by the National Key R\&D Program of China (Grants No. 2024YFA1611402, 2024YFA1611403, 2024YFA1611401, 2024YFA1611404); the National Natural Science Foundation of China (Grants No. 12393851--12393854, 12173039, 12205314, 12105301, 12305120, 12261160362, 12105294, U1931201, 12375107 and 12475114); the Department of Science and Technology of Sichuan Province (Grant No. 24NSFSC2319, 2024ZYD0111); the Project for Young Scientists in Basic Research of the Chinese Academy of Sciences (Grant No. YSBR-061); and, in Thailand, by the NSRF via the Research and Innovation Acceleration Agency for Competitiveness and Area Development (RCAD) (Program Management Unit for Technology and Innovation for Future Industries (PMU-B): Brainpower for Future Industries) (Grant No. B39G690003).

\section*{Author contributions}
All authors contributed equally to this work. In particular, S.C.~Hu, M.~Hasan (supervised by Z.G.~Yao) and Z.G.~Yao carried out the LHAASO data analysis, while Y.~Xing performed the analysis of the \fermi data. B.T.~Zhang and D.~Khangulyan contributed to the theoretical modeling and physical interpretation. R.~Wang, H.M.~Zhang and M.~Zha also contributed to the data analysis through independent cross-checks. All other authors participated in various aspects of the project, including data acquisition, processing, and quality assessment, detector calibration, event reconstruction, and simulations, and provided comments on the manuscript.

\noindent\textbf{Correspondence should be addressed to:}

Mariam Hasan \href{mailto:mariamhasan@ihep.ac.cn}{(mariamhasan@ihep.ac.cn)},

Shicong Hu
\href{mailto:hushicong@ihep.ac.cn}{(hushicong@ihep.ac.cn)},

Dmitriy Khangulyan
\href{mailto:khangulyan@ihep.ac.cn}{(khangulyan@ihep.ac.cn)},

Yi Xing
\href{mailto:yixing@shao.ac.cn}{(yixing@shao.ac.cn)},

Zhiguo Yao
\href{mailto:yaozg@ihep.ac.cn}{(yaozg@ihep.ac.cn)},

Bing Theodore Zhang
\href{mailto:zhangbing@ihep.ac.cn}{(zhangbing@ihep.ac.cn)}.

\bibliography{sample631}
\bibliographystyle{aasjournal}




\clearpage

\setcounter{figure}{0}
\setcounter{table}{0}
\setcounter{section}{0}
\setcounter{equation}{0}
\renewcommand{\thefigure}{S\arabic{figure}}
\renewcommand{\thetable}{S\arabic{table}}
\renewcommand{\thesection}{S\arabic{section}}
\renewcommand{\theequation}{S\arabic{equation}}

\renewcommand{\theHfigure}{S.\arabic{figure}}
\renewcommand{\theHtable}{S.\arabic{table}}
\renewcommand{\theHsection}{S.\arabic{section}}
\renewcommand{\theHequation}{S.\arabic{equation}}

\section*{{\large Supplementary materials}}
\section{LHAASO data and simulations}

The data utilized in this analysis were gathered by the complete LHAASO-WCDA array from March 5, 2021, to July 31, 2025, spanning a total live time of 1484 days. With the recent updates to the Cod/version of the WCDA data, which offer a significant enhancement in angular resolution at lower energy bands and a 20\% to 60\% boost in the significance of the Crab Nebula, we have opted to employ the latest version of this data for the analysis presented here.

The event selection conditions include: 
\begin{itemize}
    \item The number of triggered detector units $N_{\textrm{hit}}$, which is selected as a shower energy estimator \citep{LHAASO:2021ozi}, is divided into seven segments: $60\le N_{\textrm{hit}}<100, 100\le N_{\textrm{hit}}<200, 200\le N_{\textrm{hit}}<300, 300\le N_\textrm{hit}<500, 500\le N_\textrm{hit}<700, 700\le N_\textrm{hit}<1000$ and $1000\le N_\textrm{hit}\le 2000$;
    \item The reconstructed zenith angle is less than 50°;
    \item The parameter $P_\textrm{c}$ (defined as ``Pincness'' in~\cite{Abeysekara:2017ukg}), in which both $<\zeta_{\textrm{i}}>$ and $\sigma_{\zeta_{\textrm{i}}}$ (the mean and RMS of the the logarithmic signal amplitude distribution along distance from the shower core) are extracted from a sample set of gamma-like events, is applied to exclude hadronic showers. $P_{\textrm{c}}$ cut of six $N_{\textrm{hit}}$ segments are 1.02, 0.90, 0.88, 0.88, 0.84 and 0.84;
    \item $\rm RMDS$ is determined as root of mean distance square for shower core fitting in unit of meter based on top 10 of the hottest detectors, centered around the reconstructed shower core position. $\rm RMDS$ is smaller than 20.
\end{itemize}
The total number of gamma-like event after selection is $1.4\times10^{10}$. Detector simulation based on GEANT4~\footnote{\footnotesize \url{https://geant4.web.cern.ch/}} software package is well described elsewhere~\citep{LHAASO:2021ozi}. The simulation data used in this analysis are generated following the trajectory of Crab nebula, so there is no events with zenith angle less than 7.3 degree. The primary energy ranges from 1\,GeV to 1\,PeV and follows a power-law spectrum with index of -2.6.  The LHAASO-KM2A data used in this work are collected by full array from 20 July, 2021 to 31 July, 2025, with a total live time of 1438\,days. The event selection conditions are same as~\cite{Aharonian:2020iou}. The events are divided into bins with reconstructed energy width of 0.2 in logarithmic scale.

\section{Analysis method and results}\label{sec:supanalysis}
\subsection{Background estimation}
\begin{figure}[ht]
    \centering
    \includegraphics[width=0.6\linewidth]{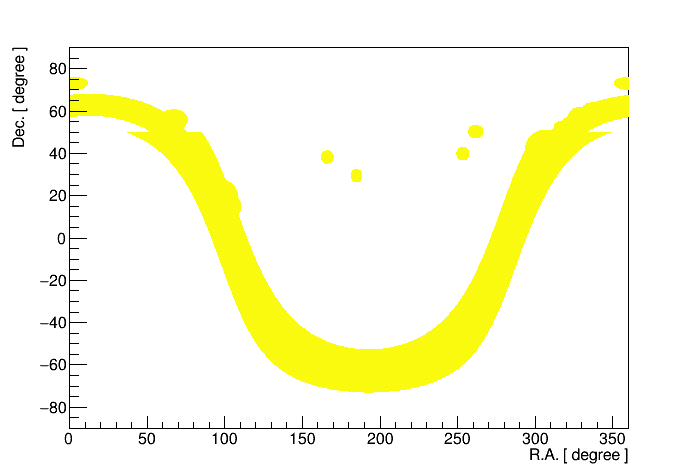}
    \caption{Masked region applied in estimating background.}
    \label{fig:1_masked_region}
\end{figure}

The celestial region from $0^{\circ}$ to $360^{\circ}$ in the right ascension (RA) and from $-20^{\circ}$ to $80^{\circ}$ in declination (Dec) is binned into cells with a size of $0.1^{\circ}\times0.1^{\circ}$, and we fill both the detected events and estimated background in those cells. Direct integral method \citep{Fleysher:2003nh} is adopted to estimate cosmic ray background. A sliding time window of 10 hours is used to estimate acceptance over the middle 4 hours of the window. As shown in Fig.~\ref{fig:1_masked_region}, the galactic plane are masked to eliminate influence of signal on background estimation. Thirteen bright sources that are far away from the galactic plane or have a large extension, with significance greater than $5\,\sigma$, are also masked off. The central position and radius of masked regions around each source are listed in Table~\ref{table1:masked_sources}. 

\begin{table*}[!htp]
\tabcolsep=0.1cm
\caption{Central position and radius of masked region around each masked sources.}
\label{table1:masked_sources}
\begin{center}
\begin{tabular*}{\textwidth}{@{\extracolsep{\fill} } cccccccccccccc}
\toprule
         RA ($^\circ$) & 339.14 & 337.07 & 1.33 & 166.11 & 253.47 & 184.91 & 261.71 & 317.08 & 330.76 & 66.99 & 105.04 & 98.12 & 307.17 \\
         Dec ($^\circ$) & 58.97  & 60.91 & 72.99 & 38.21  & 39.76  & 29.45  &  50.26 & 52.05  & 56.91  & 55.62 & 14.84  & 17.37 & 41.01  \\
         Radius ($^\circ$) & 3 & 3 & 3 & 3 & 3 & 3 & 3 & 5 & 5 & 5 & 5 & 10 & 10\\
\bottomrule
\end{tabular*}\\
\end{center}
\end{table*}

\hspace*{\fill}

\subsection{Detector response}
Detector efficiency $\eta(E, \theta)$ here is function of primary energy $E$ and incident zenith angle $\theta$. The possible dependence of the $\eta$ on the incident azimuth angle is ignored. 
\begin{equation}
    \begin{aligned}
         \eta(E, \theta) = \frac{N_{\rm{left}}(E, \theta)}{N_{\rm{all}}(E, \theta)} ,
    \end{aligned}
\end{equation}
where $N_{\rm{all}}(E, \theta)$ is the total number of simulated events, $N_{\rm{left}}(E, \theta)$ is the number of events after selection. After dividing the primary energy and zenith angle into bin, the detection efficiency maps 
are obtained. The bin width of $\theta$ and $\log_{10}(E/\rm{GeV})$ are $1^\circ$ and 0.1. Detection efficiency of bin with zenith angle smaller than $7^\circ$ is equal to the efficiency of closest bin.

Point spread function (PSF) derived from data around Crab nebula is used in this analysis. Sum of two gaussian component is adopted to describe the PSF of WCDA. The PSF parameters of six $N_{\rm{hit}}$ segments are listed in Table~\ref{table2:PSF}.
\begin{equation}
    \begin{aligned}
         f(r)=\frac{\mathcal{P}}{2\pi \sigma_{1}^{2}}\times \textrm{exp}\left(-\frac{r^{2}}{2\sigma_{1}^{2}}\right) +\frac{1-\mathcal{P}}{2\pi \sigma_{2}^{2}} \times \textrm{exp}\left(-\frac{r^{2}}{2\sigma_{2}^{2}}\right) .
    \end{aligned}
\end{equation}

\begin{table*}[!htp]
\caption{PSF of six $N_{\rm{hit}}$ segments. Where $\mathcal{P}$ is the fractional contribution from the wider Gaussian component.}
\label{table2:PSF}
\begin{center}
\begin{tabular*}{\textwidth}{@{\extracolsep{\fill} } cccccccc}
\toprule
         $N_{\rm{hit}}$ & 60--100 & 100--200 & 200--300 & 300--500 & 500--700 & 700--1000 & 1000--2000 \\
         $\mathcal{P}$ & 0.479 & 0.361 & 0.224 & 0.235  & 0.098 & 0.103 & 0.237\\
         $\sigma_{1}$ ($^\circ$) & 0.620 & 0.488 & 0.476 & 0.355 & 0.567 & 0.340 & 0.075\\
         $\sigma_{2}$ ($^\circ$) & 0.311 & 0.263 & 0.228 & 0.191 & 0.168 & 0.131 & 0.133\\
         $\phi_{68}$ ($^\circ$) & 0.672 & 0.501 & 0.404 & 0.333 & 0.277 & 0.215 & 0.181\\
\bottomrule
\end{tabular*}\\
\end{center}
\end{table*}

\subsection{Search for the target source}

The region of interest (ROI) is defined as a rectangular field centered at the position of \grs, covering the right ascension from $286.6^\circ$ to $290.6^\circ$ and the declination from $9^\circ$ to $13^\circ$. This region encompasses \grs\ along with several nearby TeV candidates, providing a sufficient margin for the spread of the spread function (PSF) and stable local background modeling.

In this analysis, no masking was applied to bright neighboring sources. Instead, their contributions were explicitly included in the likelihood model to avoid potential bias from partial masking and to achieve a self-consistent description of the overlapping emission structures within the ROI.

Based on the iterative binned Poisson maximum-likelihood method, four $\gamma$-ray sources are resolved in the ROI: three spatially extended sources (LHAASO~J1912+1010, LHAASO~J1915+1498 and \grslhaaso) and a point-like component (LHAASO~J1907+0913). The best-fit centroids, spectral parameters, and extensions (Gaussian $\sigma$; $\sigma=0$ implies point-like) obtained from the joint fit are summarized in Table~\ref{tab:roi_sources}.

\begin{figure*}[htp!]
\centering
\includegraphics[width=0.32\linewidth]{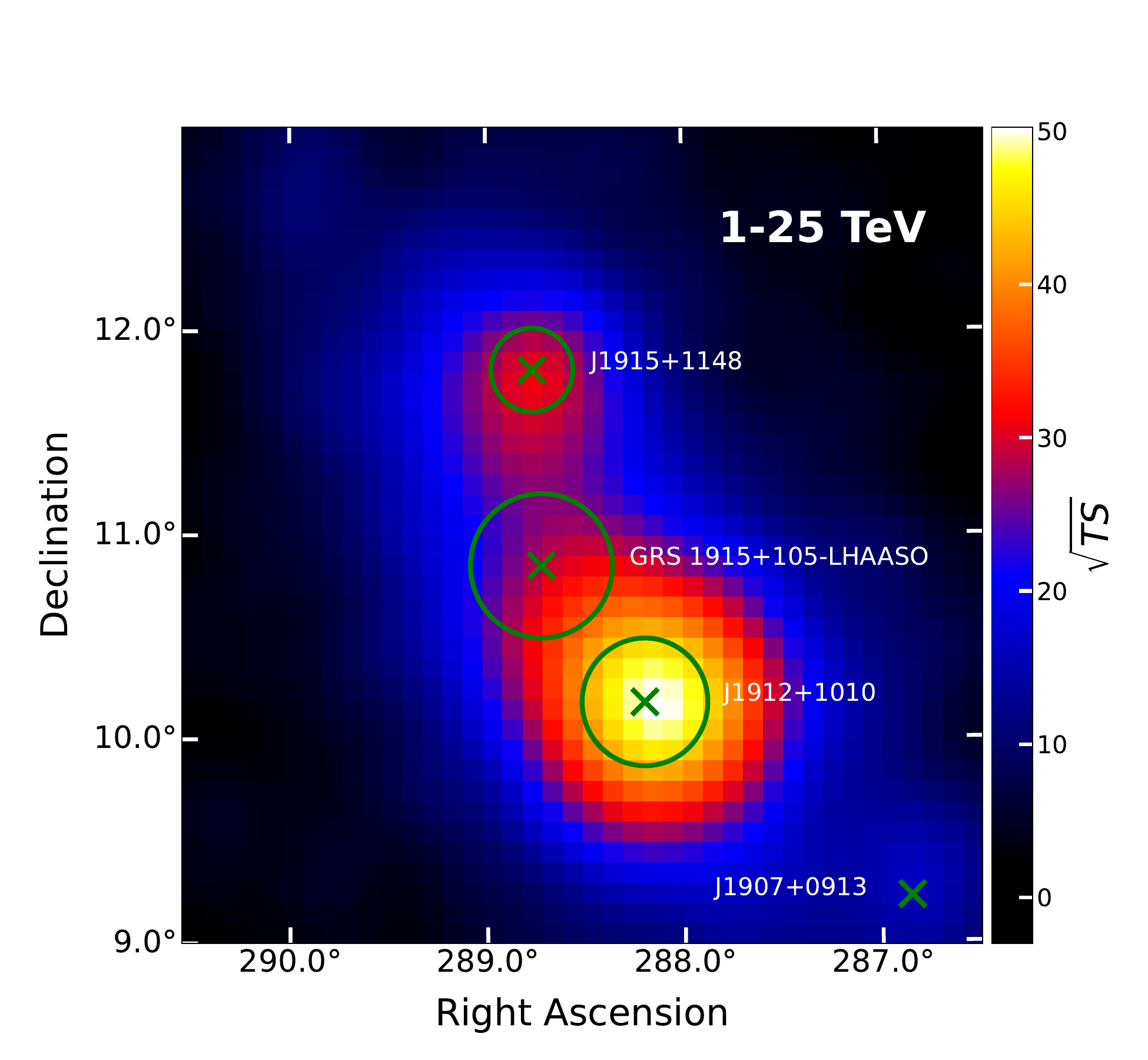}
\includegraphics[width=0.32\linewidth]{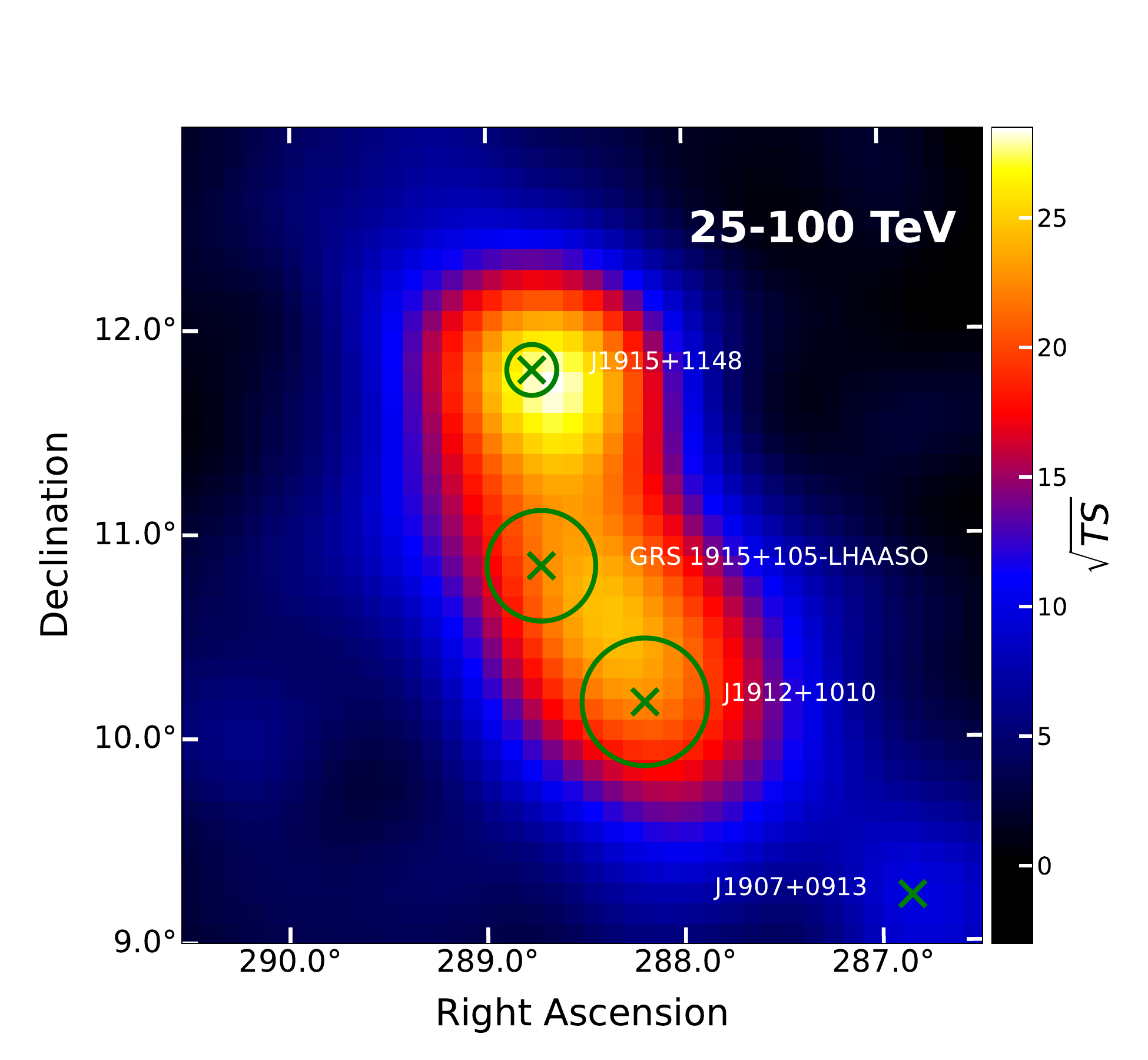}
\includegraphics[width=0.32\linewidth]{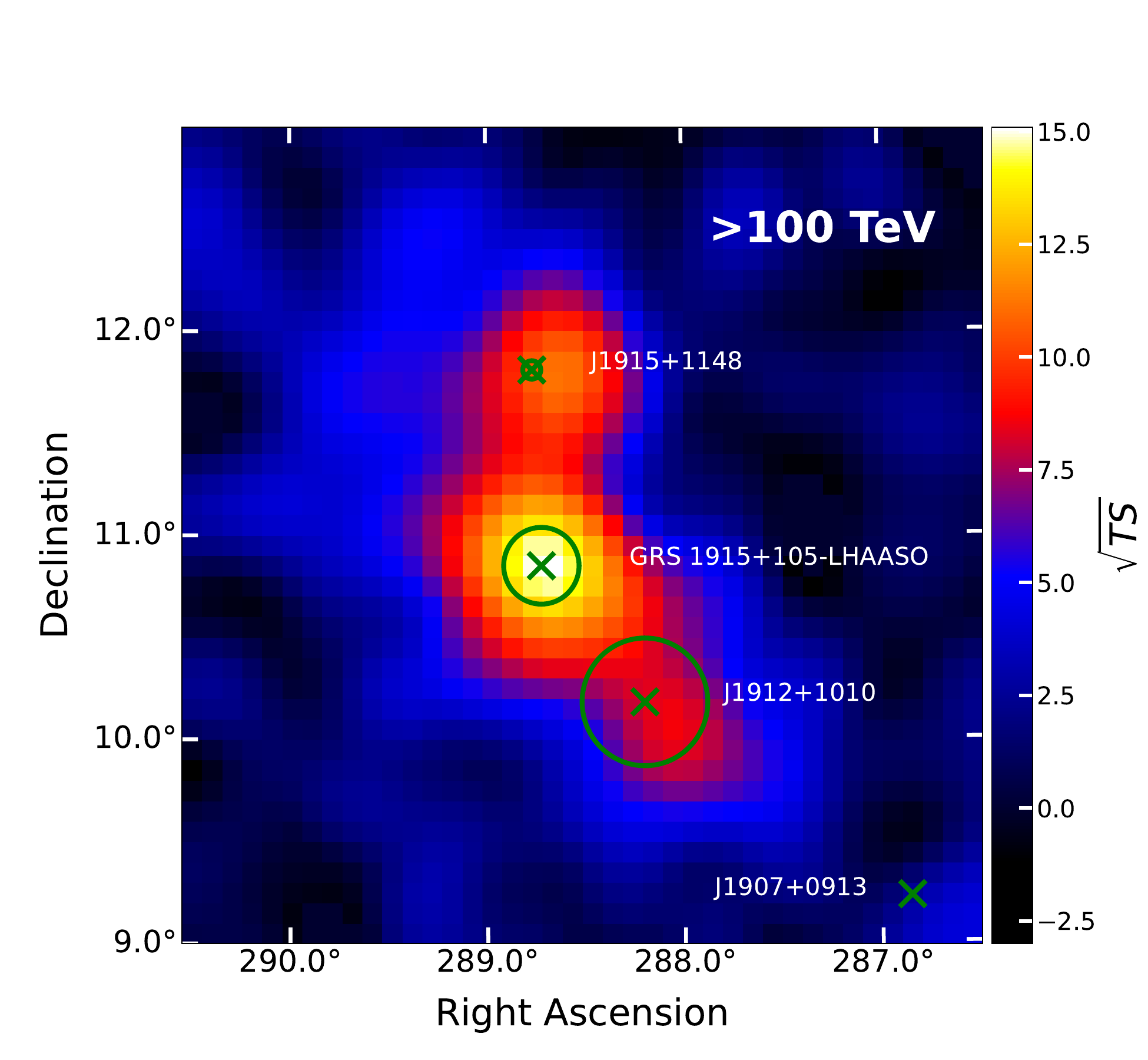}
\caption{LHAASO data significance maps of the $2^{\circ}\times2^{\circ}$ region around \grs for the full observation period. The green cross mark position of the four \gr sources; the green circles show measured extension. Panels correspond to different analysis bands (left to right).}
\label{DataSigMap}
\end{figure*}

\begin{figure*}[htp!]
\centering
\includegraphics[width=0.32\linewidth]{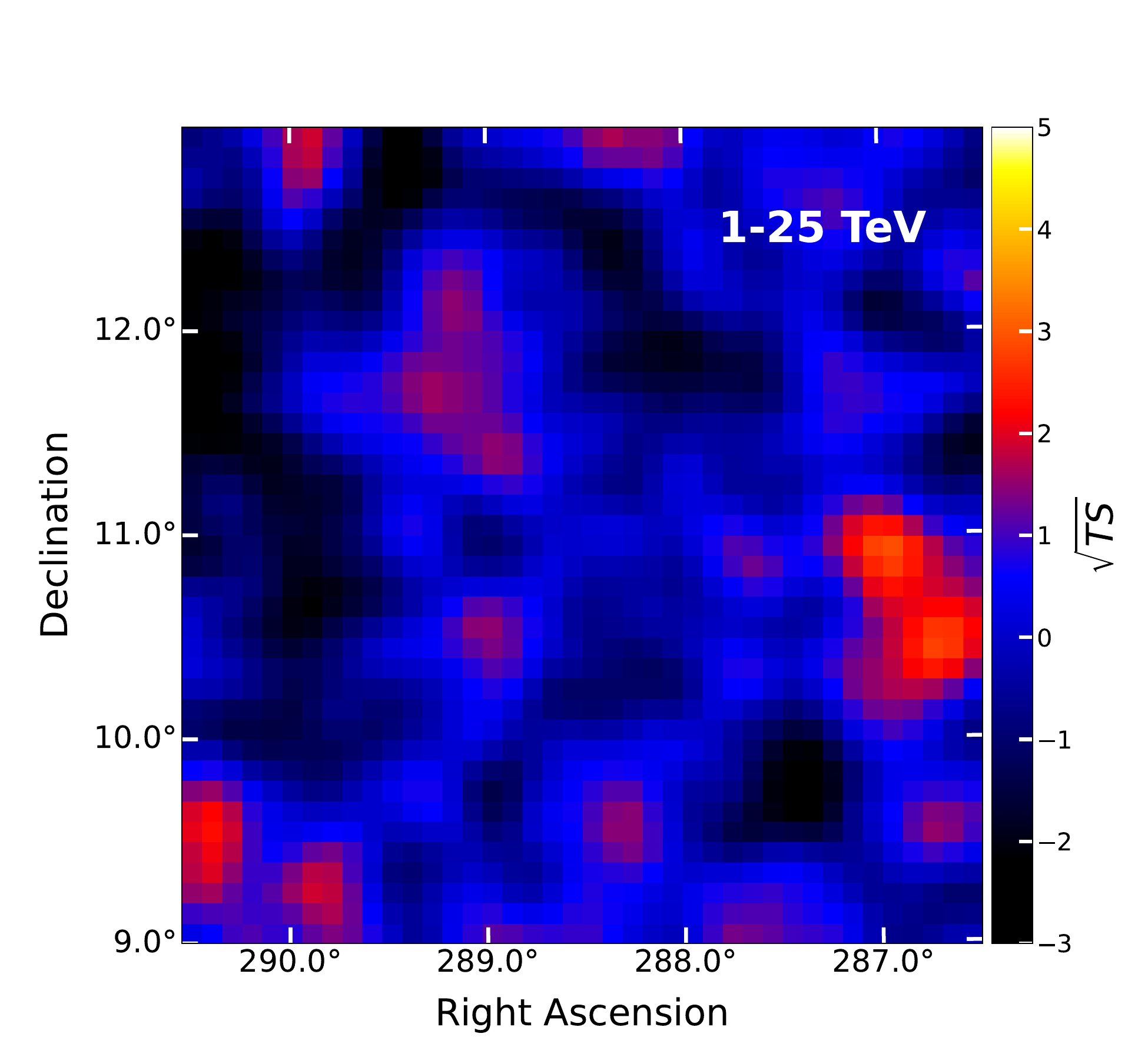}
\includegraphics[width=0.32\linewidth]{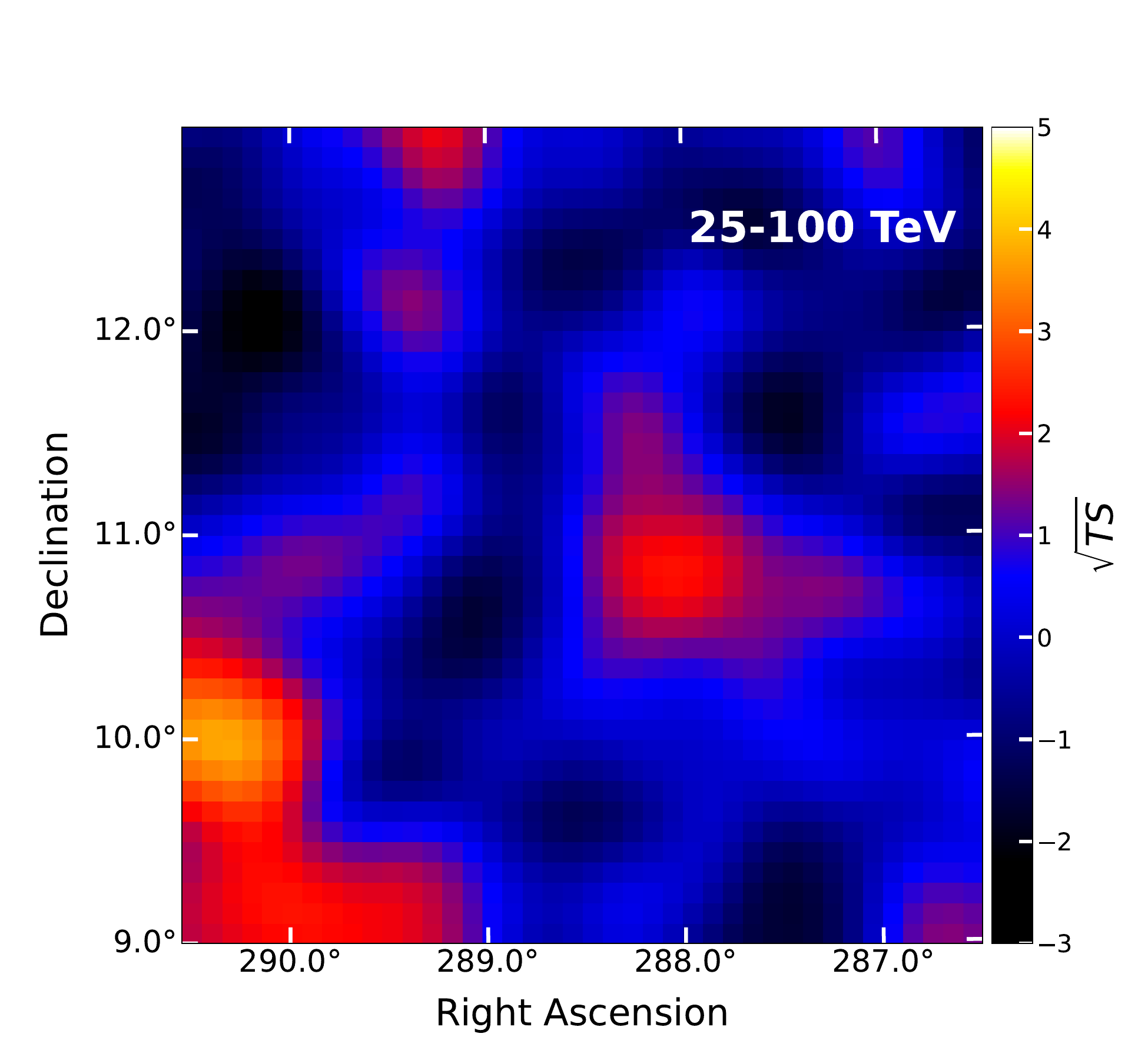}
\includegraphics[width=0.32\linewidth]{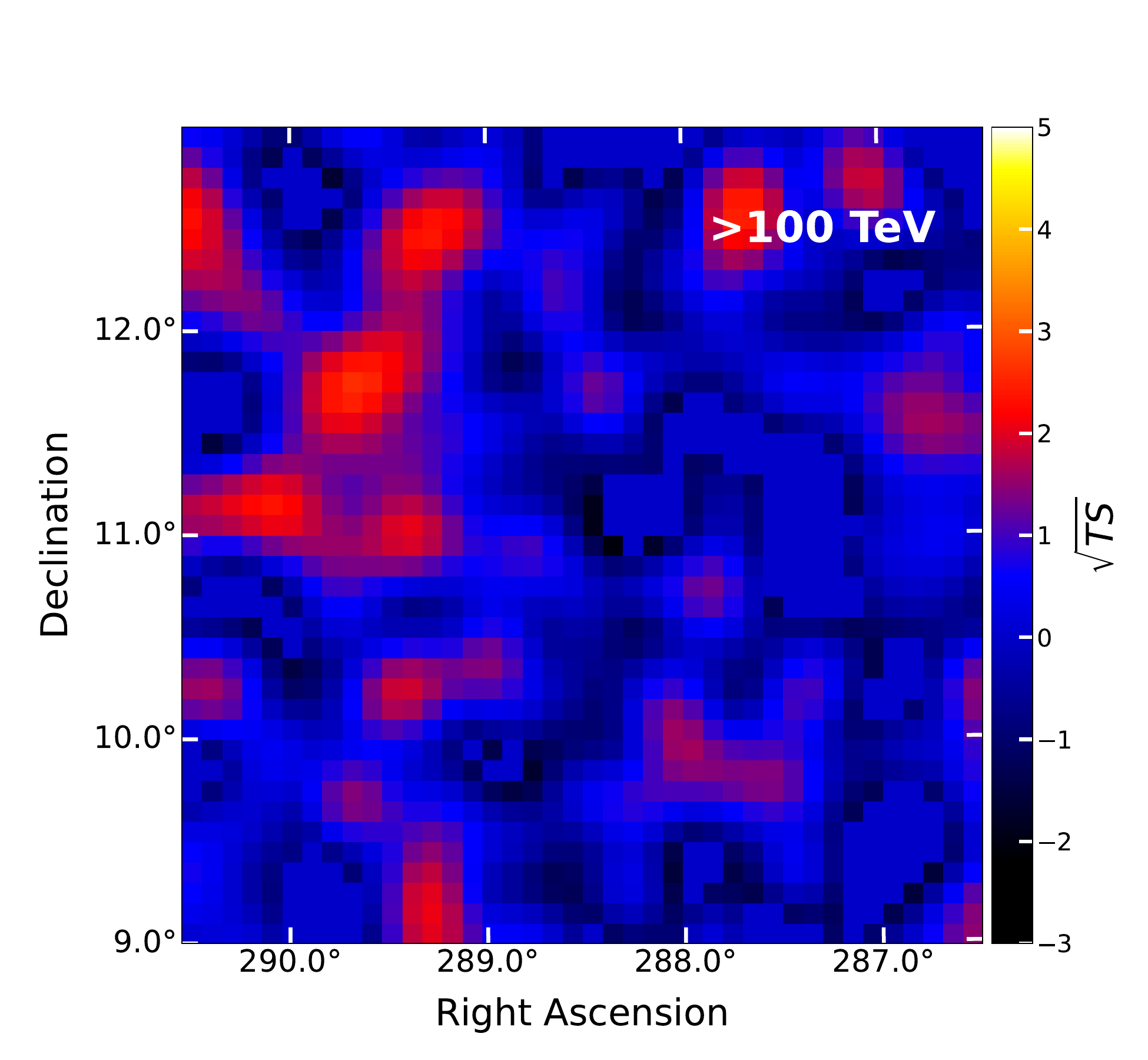}
\caption{LHAASO residual significance maps of the $2^{\circ}\times2^{\circ}$ region around \grs for the full observation period. Panels correspond to different analysis bands (left to right).}
\label{ResidualSigMap}
\end{figure*}

Fig.~\ref{DataSigMap} presents full-time data $\sqrt{TS}$ maps in a $2^\circ\times2^\circ$ field centered on \grs, illustrating the spatial distribution of excess across representative analysis bands. The corresponding residual $\sqrt{TS}$ maps are shown in Fig.~\ref{ResidualSigMap}.

\begin{table*}[!hbtp]
\caption{Sources resolved in the ROI. The source \grsobs is assigned with a new name \grslhaaso (see main text). Centroids and extensions are obtained from the joint fit, with statistical uncertainties quoted at the $1\,\sigma$ level. Sources without reported extensions are treated as point sources, while those with extensions but without a $b$ parameter are modeled using a static Gaussian template.} 

\label{tab:roi_sources}
\begin{center}
\renewcommand{\arraystretch}{1.5}
\begin{tabular}{lccccc}
\toprule
Name & RA         & Dec        & Extension  & $b$          & significance \\
     & ($^\circ$) & ($^\circ$) & ($^\circ$) & ($^\circ$) & ($\sigma$)   \\
\midrule
LHAASO J1912+1010 & $288.200^{+0.016}_{-0.016}$ & $10.176^{+0.019}_{-0.019}$ & $0.313^{+0.015}_{-0.014}$ & --  & 41.3\\
LHAASO J1915+1148 & $288.766^{+0.016}_{-0.016}$ & $11.806^{+0.017}_{-0.018}$ & $0.181^{+0.022}_{-0.022}$ & $-(0.080^{+0.040}_{-0.045})$  & 26.7  \\
LHAASO J1907+0913 & $286.851^{+0.030}_{-0.031}$ & $9.225^{+0.039}_{-0.039}$  & --  & -- & 10.6 \\
\grsobs           & $288.722^{+0.036}_{-0.036}$ & $10.846^{+0.037}_{-0.036}$ & $0.329^{+0.049}_{-0.051}$ & $-(0.083^{+0.051}_{-0.055})$  & 17.7 \\
\bottomrule
\end{tabular}
\end{center}
\end{table*}

\subsection{Galactic diffuse emission} 
\label{subsec:GDE_systematic}

\begin{figure}[ht]
    \centering
    \includegraphics[width=0.8\linewidth]{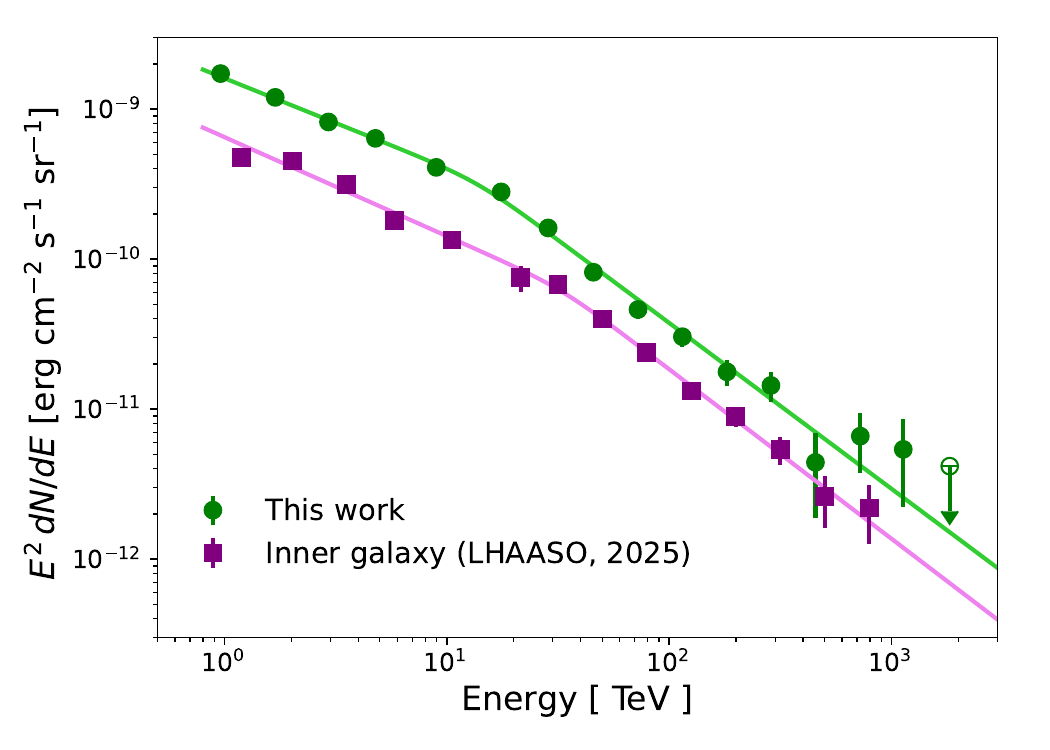}
    \caption{The best-fit SED of GDE in the enlarged ROI. The results from LHAASO diffuse paper \citep{2025PhRvL.134h1002C} is also shown as a comparison.}
    \label{fig:GDE_SED}
\end{figure}

The Galactic diffuse emission (GDE) constitutes a structured background across the ROI and can bias both spectral and morphological inferences if not modeled jointly. We therefore refit the field with an expanded Galactic–coordinate ROI,
$l\in[42.640^\circ,\,48.006^\circ]$ and $b\in[-5^\circ,\,5^\circ]$, to better constrain the GDE locally.
In this stage, the GDE template is multiplied by a power–law in energy,
\[
I_{\rm GDE}(E,l,b)\;=F_0 \left(E/E_\textrm{p}\right)^{-\alpha} \left[1 + \left(E/E_\textrm{b}\right)^{\omega}\right]^{(\alpha-\beta)/\omega}\,T(l,b),
\]
and we allow both the normalization $F_{0}$, slope $\alpha$ and $\beta$ and break energy $E_{\textrm{b}}$ to float together with source parameters. After obtaining the best–fit GDE parameters in the enlarged ROI, we freeze these values and re–optimize the source model in the nominal (smaller) ROI used throughout the analysis. This two–step procedure mitigates leakage between source flux and large–scale diffuse structure. The best-fit SED of GDE is shown in Fig.~\ref{fig:GDE_SED} and the average SED of the inner galaxy \citep{2025PhRvL.134h1002C} are also presented as a comparison.

\subsection{Absorption of CMB and interstellar radiation field}
\label{subsec:CMB_IR}

Ultra–high–energy \grn may be attenuated en route to the observer via $\gamma\gamma\! \to e^{\pm}$ interactions with ambient photon fields. For a Galactic source at $\sim 10~$kpc distances such as \grs, the relevant targets are the $T\simeq2.7$\,K cosmic microwave background (CMB) and the Galactic interstellar radiation field (ISRF: infrared dust and optical/near–IR starlight). The observed spectrum is related to the intrinsic spectrum by
\[
F_{\rm obs}(E) \;=\; F_{\rm int}(E)\,\exp\left[-\tau(E)\right],
\]
where $\tau(E)$ is the line-of-sight optical depth derived from the radiation-field model. In practice, we apply attenuation to the best-fit intrinsic spectrum (LP form from the joint fit) and refit the data with the attenuated model.

At TeV energies, ISRF absorption is usually small for typical Galactic sightlines and CMB attenuation becomes relevant mainly above 100\,TeV. Consistent with this expectation, introducing CMB and ISRF absorption produces only modest shifts in the fitted parameters and leaves the TS essentially unchanged. Table~\ref{tab:absorption} compares the fits without and with CMB and ISRF attenuation; positions and morphology are stable, while the curvature parameter $\beta$ decreases slightly, reflecting a mild degeneracy between intrinsic curvature and weak attenuation at the highest energies.

\begin{table*}[!hbtp]
\caption{Impact of IR / CMB attenuation on the LHAASO fit for \grslhaaso. Uncertainties are statistical ($1\,\sigma$).}
\label{tab:absorption}
\begin{center}
\renewcommand{\arraystretch}{1.5}
\begin{tabular}{lcc}
\toprule
Parameter & Observed & Intrinsic \\
\midrule
RA ($^\circ$)                                   & $288.718^{+0.037}_{-0.036}$ & $288.722^{+0.036}_{-0.036}$ \\
Dec ($^\circ$)                                   & $10.844^{+0.038}_{-0.037}$  & $10.846^{+0.037}_{-0.036}$ \\
$\sigma_\textrm{ext,0}$ ($^\circ$)                       & $0.328^{+0.048}_{-0.050}$   & $0.329^{+0.049}_{-0.051}$ \\
$b$ ($^{\circ}$)                                  & $-\left(0.073^{+0.047}_{-0.055}\right)$ & $-\left(0.083^{+0.051}_{-0.055}\right)$                \\
$N_0$ ($10^{-15}$ TeV$^{-1}$\,cm$^{-2}$\,s$^{-1}$) & $3.29^{+0.72}_{-0.68}$   & $3.09^{+0.70}_{-0.65}$ \\
$\alpha$                                          & $2.39^{+0.10}_{-0.13}$   & $2.39^{+0.10}_{-0.12}$ \\
$\beta$                                           & $0.085^{+0.035}_{-0.031}$   & $0.052^{+0.036}_{-0.031}$ \\
TS                                                & 9730.8           & 9730.1 \\
\bottomrule
\end{tabular}
\end{center}
\end{table*}

\paragraph{Absorption-corrected SED}
Fig.~\ref{SED} shows the joint WCDA + KM2A spectrum refit with the LP model after applying CMB (and nominal IR) attenuation along the \grs line of sight. As anticipated from Table~\ref{tab:absorption}, only mild changes occur at the highest energies; the overall curvature remains weak and the fit quality is unchanged. 

\paragraph{Implication}
The negligible change in TS and the stability of spatial parameters indicate that line-of-sight attenuation plays a sub–dominant role for this source in the WCDA + KM2A energy range. Intrinsic curvature (LP) remains preferred after accounting for plausible CMB/IR absorption, and the fitted parameter shifts are propagated as a small systematic uncertainty in the final SED.

\subsection{The choice of spectral and morphology shapes}\label{subsec:sed_mor}

\paragraph{Spectral modeling}

We fitted the joint WCDA + KM2A data with four commonly used intrinsic SED parameterizations, each referenced to $E_0 = 10~\mathrm{TeV}$:
(i) a power law (PL), $F_{\rm int}(E) = N_0\,(E/E_0)^{-\gamma}$;
(ii) a log–parabola (LP), $F_{\rm int}(E) = N_0\,(E/E_0)^{-\gamma-\beta\,\log(E/E_0)}$;
(iii) a power law with an exponential cutoff (PLEC), $F_{\rm int}(E) = N_0\,(E/E_0)^{-\gamma}\exp(-E/E_{\rm cut})$;
and (iv) smooth broken power-law (SBPL), $F_{\rm int}(E) = N_0\,(E/E_0)^{-\gamma}\left[\frac{1}{2}\left(E/E_\textrm{b}\right)^{-\omega\Delta}+\frac{1}{2}\left(E/E_\textrm{b}\right)^{\omega\Delta}\right]^{-1/\omega}$.
The best–fit parameters are summarized in Table~\ref{tab1:SED_test}. Since the break energy $E_{\rm b}$ of SBPL is not constrained well, we fixed it at the best-fit value.

\begin{figure}[!htp]
\centering
\includegraphics[width=0.8\linewidth]{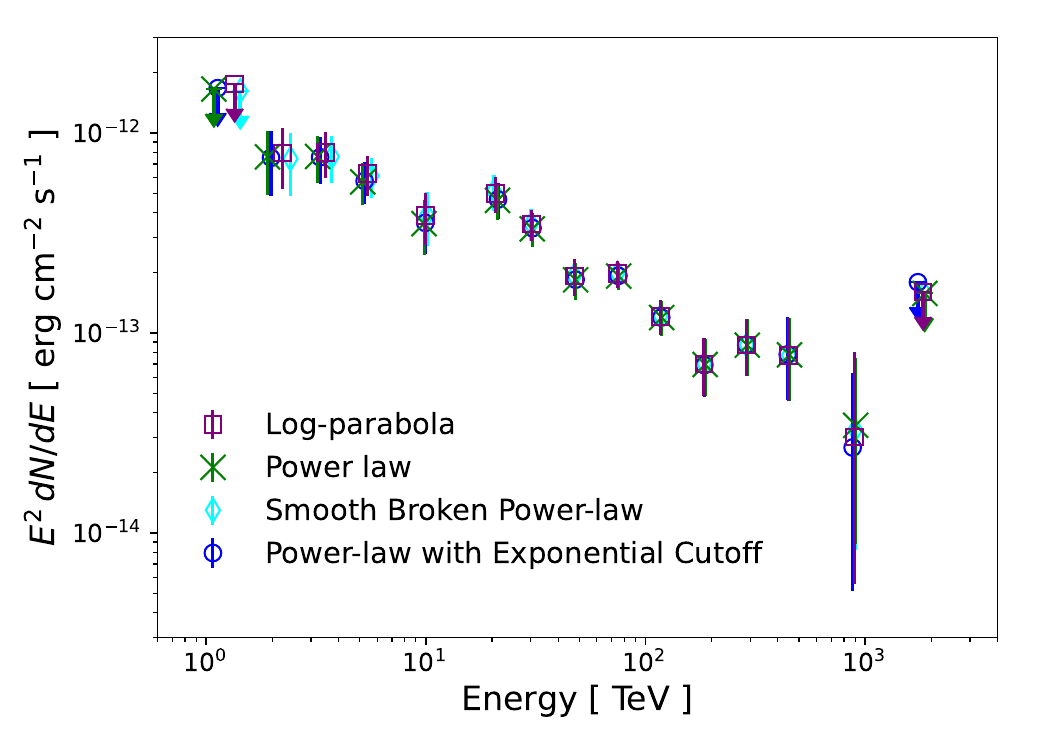}
\caption{Comparison of spectral fits to the joint (WCDA + KM2A) data. 
The four models --- power law (green), log–parabola (purplr), power law with cutoff (blue) and smooth broken power law (cyan) --- are shown with their respective $1\,\sigma$ uncertainties.
The LP provides the lowest AIC value, indicating mild but statistically significant curvature, while the PLEC gives a comparable fit without strong evidence for a cutoff.}
\label{SED_test_Joint}
\end{figure}

\begin{figure}[!htp]
\centering
\includegraphics[width=0.8\linewidth]{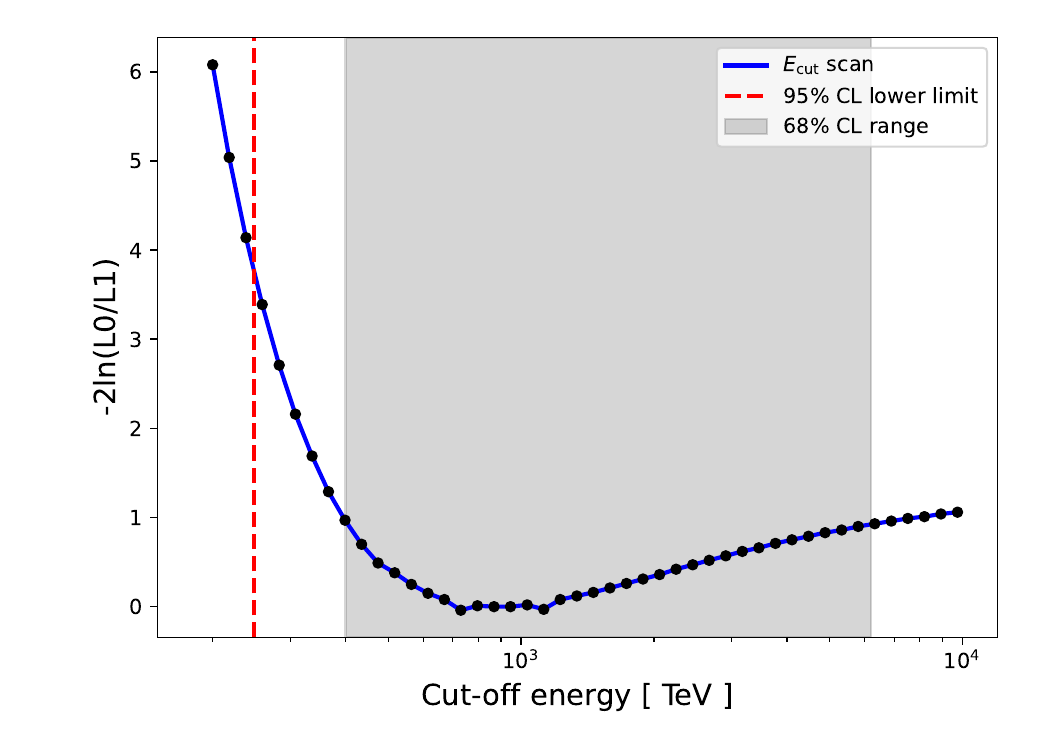}
\caption{The likelihood profile of the cut-off energy. The dashed red line is the lower limit at 95\% confidence level. The gray shaded region represents the 68.3\% confidence interval.}
\label{Ecut_scan}
\end{figure}

\begin{figure}[!htp]
\centering
\includegraphics[width=0.8\linewidth]{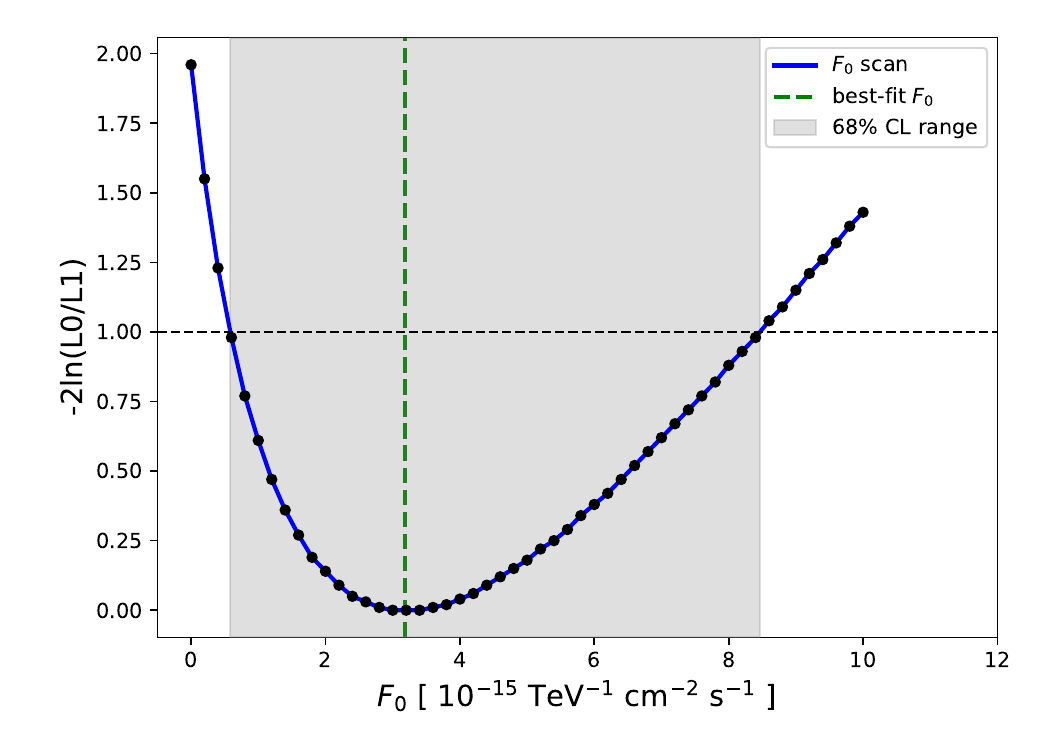}
\caption{The likelihood profile of the flux of the last measured point of intrinsic log-parabola spectrum. The dashed green line indicates the best-fit value. The gray shaded region is the 1-$\sigma$ statistical error of the flux.}
\label{PeVflux_scan}
\end{figure}

The four spectral models are compared in Fig.~\ref{SED_test_Joint}, which shows that all forms provide an adequate description of the broadband emission. The LP and PLEC curves closely trace the data across the entire range, while the simple PL tends to slightly under-predict fluxes at lower energies and over-predict at higher energies. Model comparison based on the  the Akaike Information Criterion (AIC)~\citep{AIC} indicates that the LP yields the smallest $\Delta\mathrm{AIC}$ value among the three, signifying that a mild spectral curvature is statistically preferred once model complexity is taken into account. The PLEC performs comparably in overall likelihood but does not improve the fit sufficiently to justify the additional cutoff parameter. The cut-off energy $E_{\rm cut}$ can not be constrained well. The likelihood profile method is employed to get accurate lower limit. The resulting 68.3\% confidence interval for the cutoff energy is 0.4--6.2~PeV (see Fig.~\ref{Ecut_scan}). At the 95\% confidence level, the lower limit of the cutoff energy is 248~TeV. To demonstrate the reliability of the last measurement point of the spectrum, we did profile likelihood analysis of the flux of the last point. The 68\% confidence interval is well constrained above zero as shown in Fig.~\ref{PeVflux_scan}.

\begin{table*}[!htp]
\caption{SED model fits to \grslhaaso with the LHAASO data, referenced at $E_0=10$~TeV. Uncertainties are statistical ($1\,\sigma$).}
\label{tab1:SED_test}
\begin{center}
\renewcommand{\arraystretch}{1.5}
\begin{tabular}{lccccc}
\toprule
SED model & $N_0$ ($10^{-15}$ TeV$^{-1}$cm$^{-2}$s$^{-1}$) & $\gamma$ & shape parameters & $\Delta$TS & $\Delta$AIC \\
\midrule
PL            & $2.61^{+0.54}_{-0.52}$ & $2.459^{+0.069}_{-0.077}$ & --                        & -- &  -- \\
PLEC & $2.70^{+0.59}_{-0.56}$ & $2.407^{+0.087}_{-0.10}$ & $E_{\rm cut} = 0.9_{-0.5}^{+5.3}$~PeV   & 1.4 & 0.6 \\
LP         & $3.09^{+0.70}_{-0.65}$ & $2.394^{+0.095}_{-0.12}$ & $\beta=0.052^{+0.036}_{-0.031}$   & 3.2 &  -1.2 \\
SBPL       & $3.32^{+0.83}_{-0.74}$          & $2.38^{+0.10}_{-0.14}$             & $\Delta=0.26^{+0.15}_{-0.12}$              & 3.8 &   0.2 \\
           &                        &                           &  $E_{\rm b, fixed} = 12\,\textrm{TeV}$   &     &       \\
\bottomrule
\end{tabular}
\end{center}
\end{table*}

\paragraph{Morphology tests}
Using the Log-Parabola (LP) spectrum as the baseline, we performed morphological fits on the joint data with four spatial models: a point source, a Gaussian, an elliptical Gaussian and an energy-dependent Gaussian.
The fitted parameters, centroid coordinates, extension radii, and test statistics (TS) are summarized in Table~\ref{tab5:Mor Test}. 

\begin{figure}[ht]
    \centering
    \includegraphics[width=0.8\linewidth]{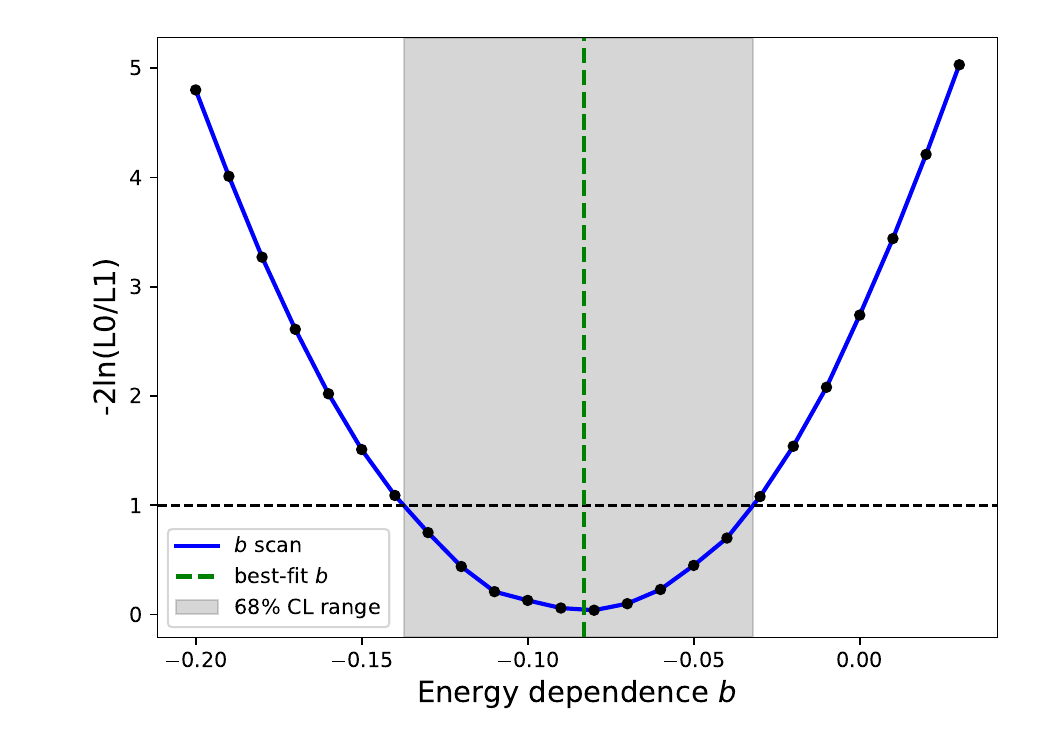}
    \caption{The likelihood profile of energy-dependence parameter $b$. The dashed green line indicates the best-fit value. The gray shaded region is the 1-$\sigma$ statistical error of $b$.}
    \label{fig:extension_b}
\end{figure}

Among the tested models, all extended morphologies improve the likelihood significantly relative to the point-source hypothesis, indicating that the emission from \grs is spatially resolved. The energy-dependent Gaussian gives the smallest $\Delta\textrm{AIC}$ and thus is prefered in our analysis. The likelihood profile of the energy-dependence parameter $b$ is shown in Fig.~\ref{fig:extension_b}. Null value 0 is out of the 68\% confidence level which indicates we did measure the energy-dependent extension. 

\begin{table*}
\caption{Morphology fits on the LHAASO data using an LP spectrum as the baseline. 
Centroid errors are statistical ($1\,\sigma$). 
$\Delta$TS values are computed relative to the point-source model.}
\label{tab5:Mor Test}
\begin{center}
\renewcommand{\arraystretch}{1.5}
\begin{tabular}{lcccc}
\toprule
Parameters & Point & Gaussian & Elliptical Gaussian & Energy-dependent Gaussian \\
\midrule
RA ($^\circ$)       & $288.748^{+0.025}_{-0.026}$ & $288.728^{+0.037}_{-0.037}$ & $288.717^{+0.040}_{-0.042}$ & $288.722^{+0.036}_{-0.036}$  \\
Dec ($^\circ$)      & $10.861^{+0.026}_{-0.026}$  & $10.848^{+0.039}_{-0.037}$  & $10.837^{+0.037}_{-0.035}$  & $10.846^{+0.037}_{-0.036}$   \\
\multirow{2}*{$\sigma$ ($^\circ$)} & \multirow{2}*{--}                 & \multirow{2}*{$0.267^{+0.038}_{-0.038}$}   & $0.312^{+0.054}_{-0.051}$   & \multirow{2}*{$0.329^{+0.049}_{-0.051}$}    \\
                    &                   &                   & $0.231^{+0.047}_{-0.045}$   &                     \\
$b$ ($^\circ$)      & --                & --                & --                 & $-(0.083^{+0.051}_{-0.055})$  \\
$\theta$ ($^\circ$) & --                 & --                & $7.3^{+23}_{-28}$        & --                   \\
$\Delta$TS          & --                 & 30.0              & 32.7              & 32.8              \\
$\Delta$AIC         & --                 & -28               & -26.7             & -28.8              \\
\bottomrule
\end{tabular}
\end{center}
\end{table*}

\paragraph{Summary}
Taken together, the SED and morphology analyses favor a gently curved log-parabola spectrum and confirm that the source is spatially extended with an energy-dependent Gaussian extension. The centroid remains stable across models, and the inferred extension does not materially affect the spectral parameters within uncertainties.

\subsection{Influence of cataloged SNR, PSR and 4FGL sources}
\label{subsec:SNR_PWN_influence}

To assess possible contamination from nearby cataloged sources, we examined all pulsars, SNRs, and GeV sources located within $0.5^\circ$ of the best-fit position of \grslhaaso. In particular, we tested SNR G045.7$-$00.4, PSR J1914+1054g, 4FGL J1916.3+1108 and 4FGL J1914.5+1107c. For each candidate, an additional extended-like source component was added at the catalog position while keeping its position and spectral shape fixed. The ROI was then refit with the GDE parameters fixed to the values derived in Section~\ref{subsec:GDE_systematic}.

The inclusion of the cataloged sources yields only marginal changes in likelihood, with $\Delta\mathrm{TS}=3.1$, $8.2$, $0.1$ and $0.0$ for SNR G045.7$-$00.4, PSR J1914+1054g, 4FGL J1916.3+1108 and 4FGL J1914.5+1107c, respectively. In all cases, the source position, extension, and spectral parameters of \grslhaaso remain consistent with the baseline fit within statistical uncertainties. These results indicate that nearby cataloged sources do not significantly contribute to the observed TeV emission and do not affect the inferred properties of \grslhaaso.

\section{\fermi-LAT data reduction}
\label{sect:app_lat}

\subsection{Observations}
\label{sec:obs}

The Large Area Telescope (LAT)
is the main instrument onboard \fermi, it is a \gr\ imaging instrument which 
can continuously
scans the whole sky in tens of MeV to hundreds of GeV band.
In the analysis, we selected the LAT events with energies of 0.1--500 GeV
from the updated \fermi Pass 8 database.
A 20$^{\circ}$ $\times$ 20$^{\circ}$ square region centered at \grslhaaso\ was defined as the region of interest (ROI).
We included the LAT events during the time period from 2008-08-04 15:43:39 (UTC) to 2025-11-26 23:19:55 (UTC).
The \textit{SOURCE} event class was used, which is 
recommended for most types of \gr\ sources, and both of the \textit{Front} and \textit{Back} event types were considered in the analysis.
We included the events with zenith angles less than 90 degrees to prevent the Earth's limb contamination, and excluded the periods when the spacecraft 
was in the South Atlantic Anomaly (SAA) or experiencing instrumental 
anomalies (recommended by the LAT team~\footnote{\footnotesize \url{http://fermi.gsfc.nasa.gov/ssc/data/analysis/scitools/}}).
The data analysis was performed using the standard \fermi\ data analysis software package \textit{Fermitools} 2.2.0.

\subsection{Source identification}
\label{sec:si}

We included all sources within 20 degrees centered at \grslhaaso\ in the source model. The source positions and the spectral parameters are provided in the \fermi-LAT 14-year source catalog \citep{4fgl-updated4}.
We set the spectral parameters of the sources 
within 4 degrees, and the normalization parameters outside of 4 degrees but within 5 degrees from \grslhaaso\ as free parameters. 
All of the other parameters of the catalog sources were fixed at their catalog values.
The spectral model gll\_iem\_v07.fit was used for the Galactic diffuse emission, and the spectral file iso\_P8R3\_SOURCE\_V3\_v1.txt was used for the extragalactic diffuse emission. The normalizations of these two diffuse components were set free. This source model is the baseline source model used in the analysis.

\begin{figure*}[!hbtp]
\centering
   \includegraphics[width=0.42\textwidth]{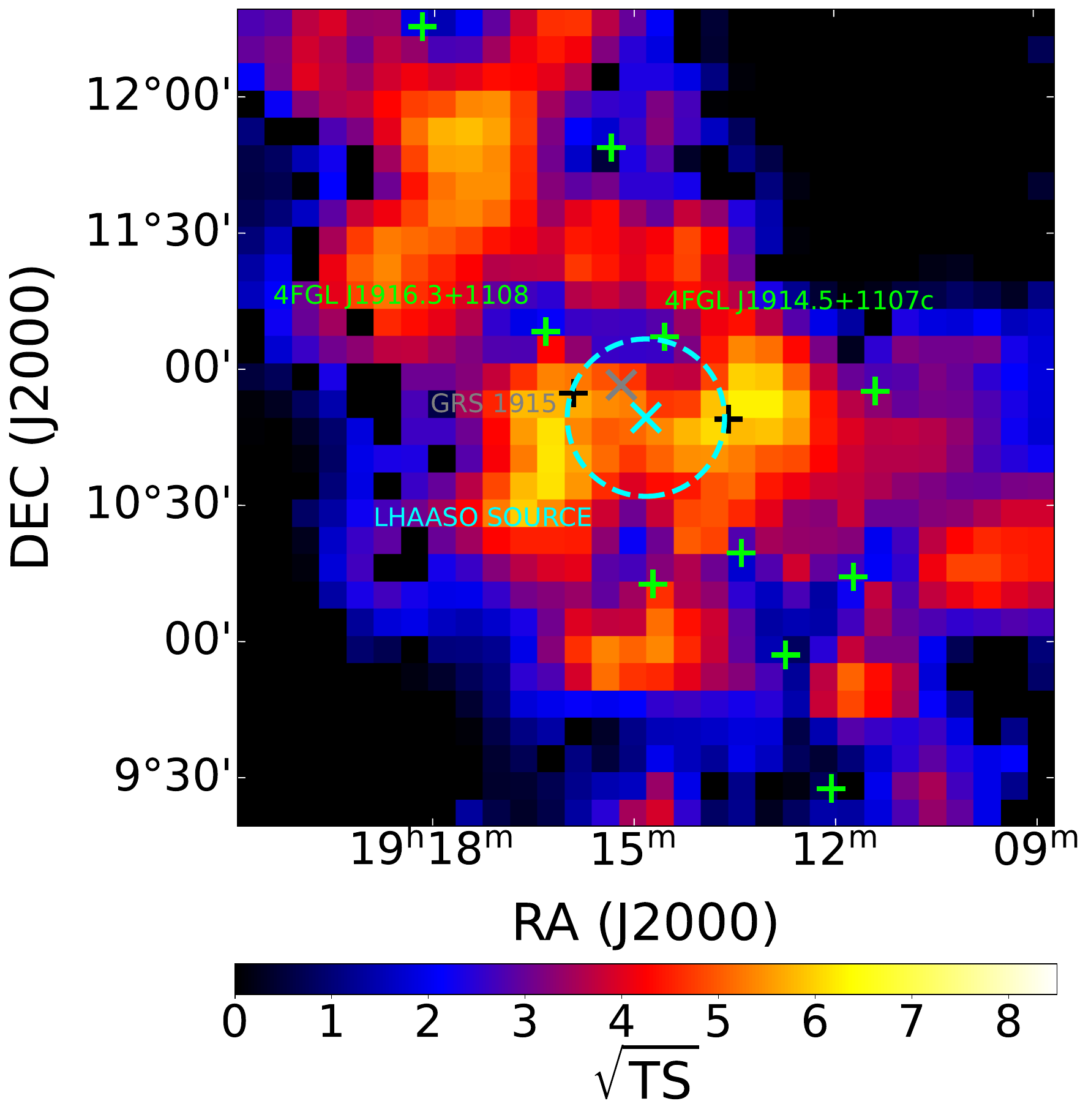}
      \includegraphics[width=0.42\textwidth]{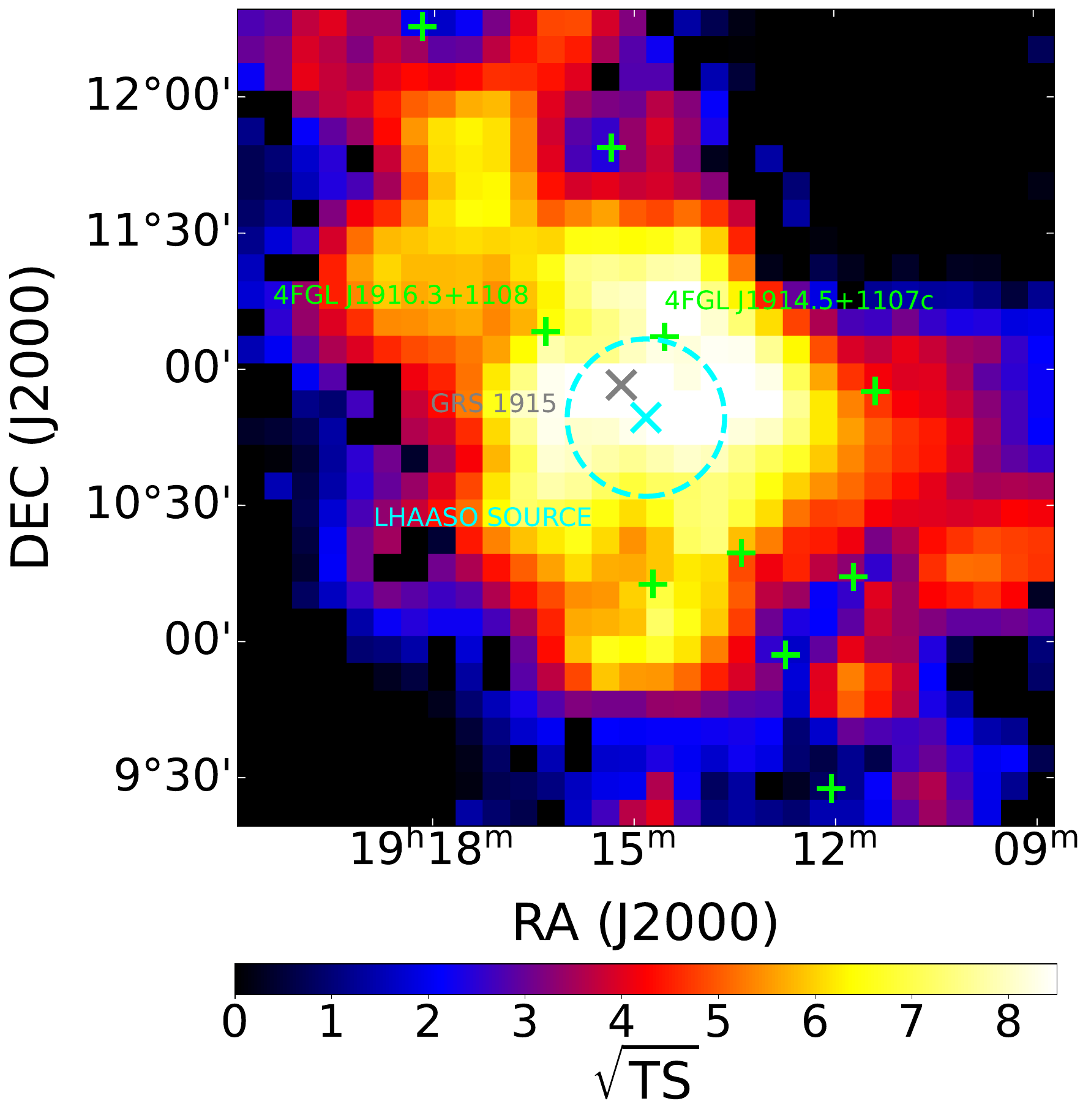}
   \caption{1--500~GeV significance maps for the region of \grslhaaso. All catalog sources were considered and removed in the {\it left} panel. \grslhaaso\ was included in the source model to perform the likelihood analysis, and all catalog sources were removed in the {\it right} panel. The green pluses mark the positions of the \fermi\ catalog sources, the gray cross marks the position of \grs, the cyan cross and dashed circle mark the center position and LHAASO-derived extension of \grslhaaso, respectively. The two additional point sources, possibly associated with PS J1915.5+1056 and PS J1913.4+1050 in \cite{2025ApJ...979L..40M}, are marked with two black pluses in the {\it left} panel (see Supplementary Section~\ref{sec:contamination}).
}
   \label{fig:tsmap}
\end{figure*}

With the \fermi\ data in the ROI and the baseline source model, we performed the standard binned likelihood analysis in the 0.1--500 GeV energy band. The resulting log-likelihood value is 31156670.8 (defined as $\ln\mathcal{L}_{0}$). Based on the fitted source model, we derived the residual significance map of a 3$^{\circ}$ $\times$ 3$^{\circ}$ region centered on \grslhaaso\ in the 1--500 GeV energy band. The significance was estimated as the square root of the TS value, and this energy range was adopted to reduce the higher contamination from diffuse emission and nearby sources in the lower energy range.
The resulting significance map is shown in the left panel of Fig.~\ref{fig:tsmap}. It can be seen that noticeable emission residuals are present around the center of the map, suggesting that the baseline source model does not fully describe the \gr\ emission in this region.

We then included the counterpart for \grslhaaso\ in the source model and re-performed the likelihood analysis in the 0.1--500~GeV energy band. The \gr\ counterpart for \grslhaaso\ was described with the template derived from the LHAASO observations. The \gr\ emission from the source was modeled with a simple power law $\dd N/\dd E=N_{0} E^{-\Gamma}$. We obtained a $\Gamma$ of 2.07$\pm$0.04 and a 0.1--500 GeV photon flux of 3.4$\pm$0.5 $\times$ 10$^{-8}$ photons cm$^{-2}$ s$^{-1}$ for \grslhaaso, with a TS value of 223. The $\ln\mathcal{L}$ value is 31156717.5 (defined as $\ln\mathcal{L}_{1}$). 

By comparing the log-likelihood values, we found that the source model including \grslhaaso\ provides a significantly better fit than that without it. The TS of the fit improvement is 93.3 (defined as 2$\times$(log$\textit{L}_{1}$ $-$ log$\textit{L}_{0}$), see Table~\ref{tab:ts}). We also tested an alternative source model by replacing the \grslhaaso\ template with a point source (PS) located at its center. No significant fit improvement was found compared to the baseline source model (see Table~\ref{tab:ts}).
Using the fitted source model including \grslhaaso, we derived the significance map with \grslhaaso\ revealed in the map and show it in the right panel of Fig.~\ref{fig:tsmap}. The emission profile seems to be consistent with that of \grslhaaso. We note that a prior LHAASO source template was adopted in the above analysis. In Supplementary Section~\ref{sec:contamination}, we carefully assess the impact of emission contamination from surrounding sources on our results.

\begin{table*}[!htp]
\caption{Comparison of Test Statistics of the fit improvement. The test statistics TS are estimated by $-2\ln(\mathcal{L}_{0}/\mathcal{L}_{1})$, where $\mathcal{L}_{0}$ was obtained with a source model without any additional sources at the GRS 1915 region (the baseline model).}
\label{tab:ts}
\begin{center}
\begin{tabular}{lccccc}
\toprule
Model & TS \\ 
\midrule
4FGL J1916.3 (PS, baseline) & 0.0 \\ 
4FGL J1916.3 (PS) $+$  \grslhaaso\ (PS) & 7.6 \\ 
4FGL J1916.3 (PS) $+$  \grslhaaso\ (EXT = 0.267$^{\circ}$) & 93.3 \\
\midrule
4FGL J1916.3 (EXT) & 98.9 \\
4FGL J1916.3 (EXT) $+$  \grslhaaso\ (PS) & 99.4 \\
4FGL J1916.3 (EXT) $+$  \grslhaaso\ (EXT = 0.267$^{\circ}$) & 140.6 \\
4FGL J1916.3 (EXT) $+$  \grslhaaso\ (EXT = 0.478$^{\circ}$) & 166.0 \\
\midrule
4FGL J1916.3 (EXT) $+$  TWO NEW PS & 122.0 \\
4FGL J1916.3 (EXT) $+$  TWO NEW PS $+$  \grslhaaso\ (EXT = 0.267$^{\circ}$) & 145.0 \\
\bottomrule
\end{tabular}
\end{center}
\end{table*}

\subsection{Spectral Analysis}
\label{sec:sa}

\begin{figure*}
   \centering
   \includegraphics[width=0.31\textwidth]{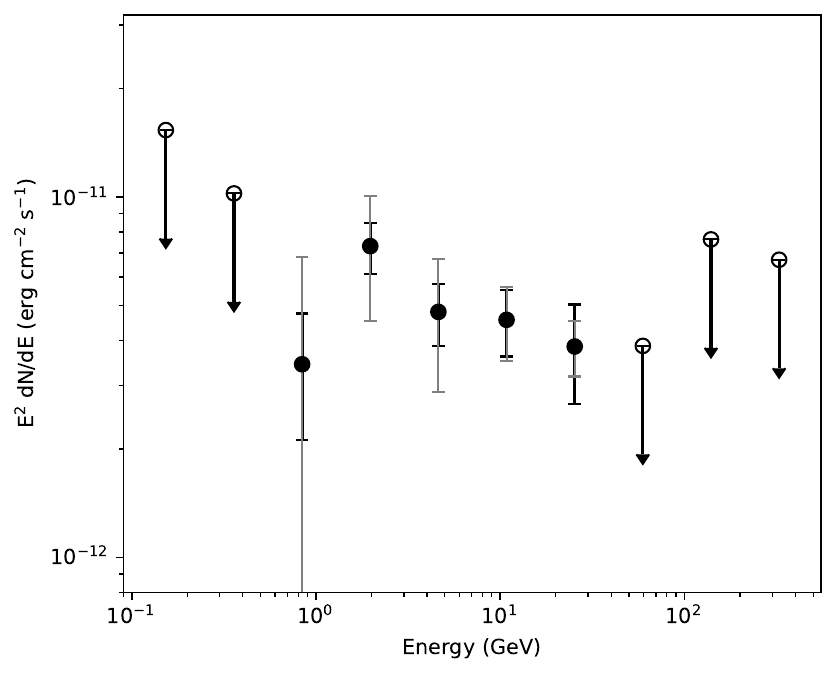}
   \includegraphics[width=0.31\textwidth]{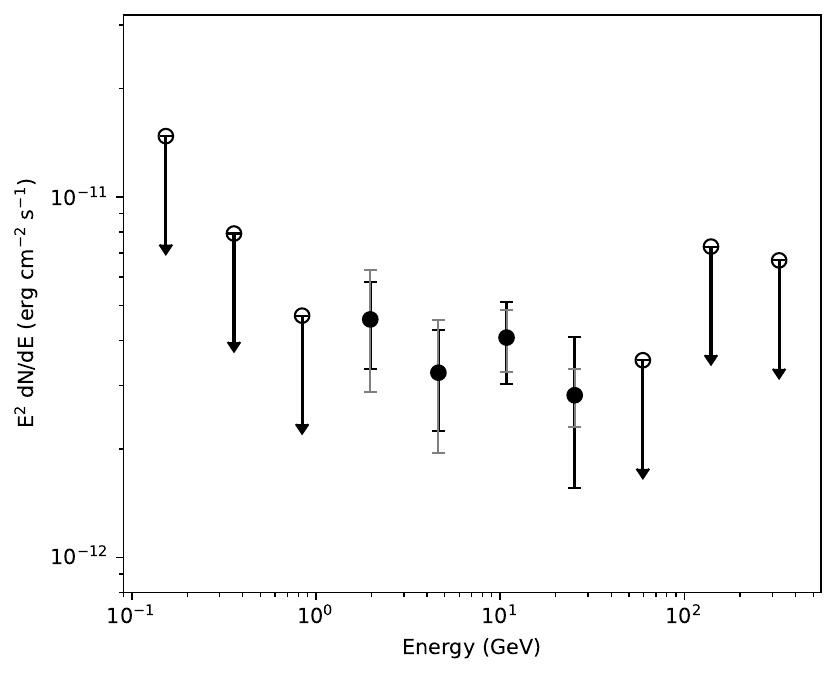}
    \includegraphics[width=0.31\textwidth]{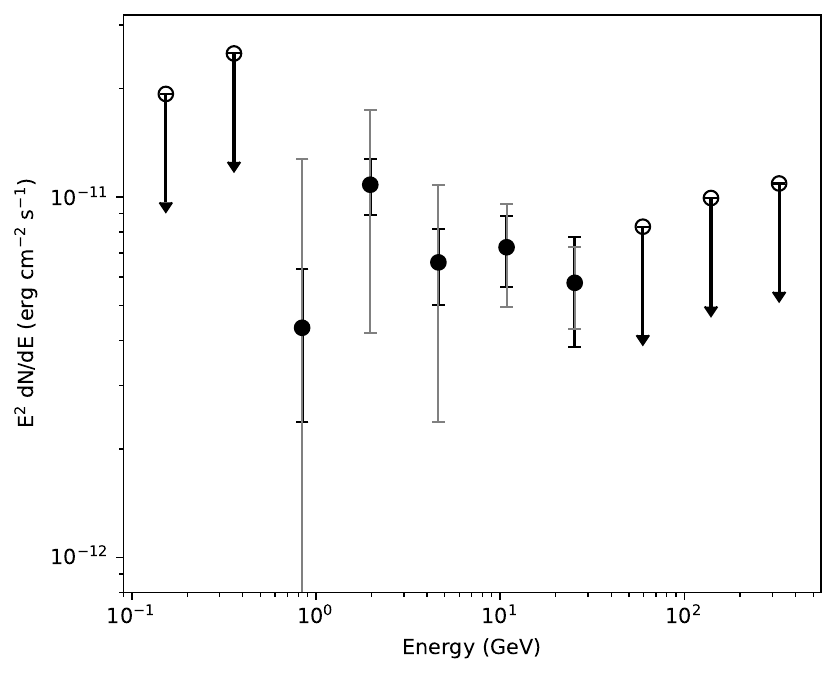}
   \caption{\gr\ spectra for \grslhaaso\ obtained with the source models 4FGL J1916.3 (PS) $+$  \grslhaaso\ (EXT 0.267$^{\circ}$) in the {\it left}  panel, 4FGL J1916.3 (EXT) $+$  \grslhaaso\ (EXT 0.267$^{\circ}$) in the {\it middle} panel, and 4FGL J1916.3 (EXT) $+$  \grslhaaso\ (EXT 0.478$^{\circ}$) in the {\it right} panel. The statistic and systematic uncertainties are plotted with black and gray error bars respectively.
   }
   \label{fig:spectra}
\end{figure*}

\begin{table*}[!htp]
\caption{Flux Measurements. The energy flux is defined as $E^{2} \dd N/\dd E$.  The first and second uncertainties are statistic and systematic uncertainties, respectively.  Energy fluxes without uncertainties are the 95\% upper limits.}
\label{tab:spectra}
\begin{center}
\begin{tabular}{cccccccc}
\toprule
 & PS + \{EXT = $0.267^{\circ}\}$ & EXT + \{EXT = $0.267^{\circ}\}$ & EXT + \{EXT = $0.478^{\circ}$\} \\
\midrule
Energy (MeV) & \multicolumn{3}{c}{Energy flux ($10^{-12}\rm\, erg\, cm^{-2}\, s^{-1}$)} \\
\midrule
153.1 & 15.34 & 14.76 & 19.33 \\
358.8 & 10.23 & 7.92  & 25.03 \\
840.9 & 3.44$\pm$1.32$\pm$3.37 & 4.69 & 4.34$\pm$1.97$\pm$8.40 \\
1970.8 & 7.31$\pm$1.19$\pm$2.78 & 4.58$\pm$1.25$\pm$1.70 & 10.82$\pm$1.90$\pm$6.63 \\
4618.9 & 4.80$\pm$0.94$\pm$1.92 & 3.26$\pm$1.01$\pm$1.31 & 6.59$\pm$1.56$\pm$4.21 \\
10825.1 & 4.57$\pm$0.95$\pm$1.06 & 4.08$\pm$1.05$\pm$0.80 & 7.26$\pm$1.63$\pm$2.30 \\
25370.6 & 3.85$\pm$1.19$\pm$0.67 & 2.82$\pm$1.26$\pm$0.51 & 5.78$\pm$1.95$\pm$1.48 \\
59460.3 & 3.87 & 3.53 & 8.28 \\
139355.0 & 7.63 & 7.29 & 9.94 \\
326604.0 & 6.70 & 6.67 & 10.91 \\
\bottomrule
\end{tabular}
\end{center}
\end{table*}

We extracted the \gr\ spectrum of \grslhaaso\ counterpart by performing the maximum likelihood analysis of the LAT data in 10 evenly 
divided energy bands in logarithm from 0.1--500 GeV. 
In the extraction, the spectral normalizations of the sources within 5 
degrees from the source position were set as free parameters, while all the other parameters 
of the sources were fixed at the values obtained from the maximum 
likelihood analysis. The \gr\ emission of \grslhaaso\ was described as a simple power law, with the $\Gamma$ values fixed to 2.
We kept only spectral data points with fluxes greater than twice their uncertainties, and derived 95\% flux upper limits 
otherwise. 

We note that the systematic uncertainties dominate the uncertainties of the spectral results in the low energy range. We considered the systematic uncertainties induced by the galactic diffuse emission by repeating the likelihood analysis in each energy band with the normalizations of the galactic diffuse component artificially fixed to the $\pm$6\% deviation from the best-fit values. This deviation represents the local departure from the best-fit diffuse model, and it was found to be $\sim$6\% when analyzing source-free regions on the galactic plane \citep{abdo+w51c2009,abdo+w28-2010}.
The obtained spectra are plotted in the left panel of Fig.~\ref{fig:spectra}, and the flux values of the spectral data points are provided in Table~\ref{tab:spectra}. 

\subsection{Variability analysis}
\label{sec:vi}

We searched for any
long-term variability of the \grslhaaso\ counterpart in 0.1--500 GeV band by calculating the variability 
indice TS$_{var}$ for the \gr\ emission with 106 time bins 
(each bin constructed from 60-day data), following the procedure 
introduced in \cite{nol+12}. 
If the flux is constant, 
TS$_{var}$ would be distributed as $\chi^{2}$ with 105 degrees of freedom. 
Variable sources would be identified with TS$_{var}$ larger than 141.6
(i.e., at a 99\% confidence level).
The computed TS$_{var}$ is 86.5, indicating that there were no significant 
long-term variability observed in the \gr\ emission. 

\subsection{Contamination from nearby sources}
\label{sec:contamination}

\grslhaaso\ lies in a crowded region containing a lot of \gr\ catalog sources. The two nearest ones are 4FGL J1914.5+1107c and 4FGL J1916.3+1108, which were detected with TS of 33 and 463, respectively. Given the much higher significance of 4FGL J1916.3+1108, we primarily considered its potential \gr\ contamination. This source was found to be the \gr\ counterpart for SNR G045.7--00.4 \citep{Zhang:2021jeq,2025ApJ...979L..40M}. In \cite{2025ApJ...979L..40M} an extension with a $\sim$0.23$^{\circ}$ radius (68\% contaminant radius) was defined for 4FGL J1916.3+1108, and a point source was detected and considered to be the \gr\ counterpart of \grs. We firstly tested three source models, which include only the extended source at 4FGL J1916.3+1108 (EXT), the extended source 4FGL J1916.3+1108 and a point source at the center of \grslhaaso\ (EXT + PS), and the extended source 4FGL J1916.3+1108 and the \grslhaaso\ counterpart (EXT + \{EXT = $0.267^{\circ}$\}). The TS of the fit improvement relative to the baseline model are listed in Table~\ref{tab:ts}. The source model including the extended source 4FGL J1916.3+1108 and the \grslhaaso\ counterpart provides the best fit with the largest TS of 140.6. Compared to the source model including only the extended source 4FGL J1916.3+1108, the fit improvement of further including \grslhaaso\ counterpart is 42. A spectral index $\Gamma = 2.00 \pm 0.05$ and a 0.1--500 GeV photon flux of $1.9 \pm 0.3 \times 10^{-8}\rm\, photons\, cm^{-2}\, s^{-1}$ were obtained for \grslhaaso\ with a TS value of 103. This spectral index is consistent with that obtained in the above analysis, while the \gr\ flux and the TS are lower, which is expected since an extended template for 4FGL J1916.3+1108 was adopted. We extracted the \gr\ spectrum of \grslhaaso\ with this source model. The obtained spectrum is plotted in the middle panel of Fig.~\ref{fig:spectra}, and the flux values of the spectral data points are provided in Table~\ref{tab:spectra}. The \gr\ spectrum has lower fluxes compared to that obtained with the source model considering 4FGL J1916.3+1108 as a point source (the left panel of Fig.~\ref{fig:spectra}), yet within the uncertainty ranges.

We noticed that the angular size for the SNR defined in~\cite{Zhang:2021jeq} is consistent with that in~\cite{2025ApJ...979L..40M}, but the center of the \gr\ counterpart is $0.12^{\circ}$ from 4FGL J1916.3+1108. We replaced the \gr\ counterpart for 4FGL J1916.3+1108 in the above source models, with the SNR gaussian template given in \cite{Zhang:2021jeq}. The source models with an extended source centered at 4FGL J1916.3+1108 always provide better fit than those centered at the position given in \cite{Zhang:2021jeq}, so throughout this paper we adopted the position of 4FGL J1916.3+1108 as the center position.

In~\cite{2025ApJ...979L..40M} they suggested that a point source PS J1915.5+1056 (marked as a black plus near \grs\ in the left panel of Fig.~\ref{fig:tsmap}) is possibly associated with \grs, we further added this new point source in our source models. In addition, we determined the position with the highest TS value in the left panel of Fig.~\ref{fig:tsmap}. An additional point source located at RA = $288.397^{\circ}$, Dec = $10.820^{\circ}$, with a 68\% error circle of $0.15^{\circ}$ was found (also marked as a black plus in the left panel of Fig.~\ref{fig:tsmap}) besides PS J1915.5+1056.
This best-fit position (RA, Dec = 19:13:35.28, 10$^\circ$49'12.0") is likely consistent with another point source PS J1913.4+1050 defined in~\cite{2025ApJ...979L..40M}, although the position was not given in their paper since they suggested that PS J1913.4+1050 is unlikely to be astrophysical origin. Nevertheless, we also added this source in our source models and performed the likelihood analysis to check whether the source model with these two additional new point sources (TWO NEW PS) or that with one extended source \grslhaaso\ provides a better fit. The TS of the fit improvement relative to the baseline model are listed in Table~\ref{tab:ts}. It can be seen that the source model including \grslhaaso\ provides a better fit than that with the two additional point sources. Furthermore, adding these two point sources does not significantly improve the fit. Therefore, the two new sources are not included in the analysis.

\subsection{Spatial distribution analysis}
\label{sec:sda}

\begin{figure}[!htp]
\centering
      \includegraphics[width=0.45\textwidth]{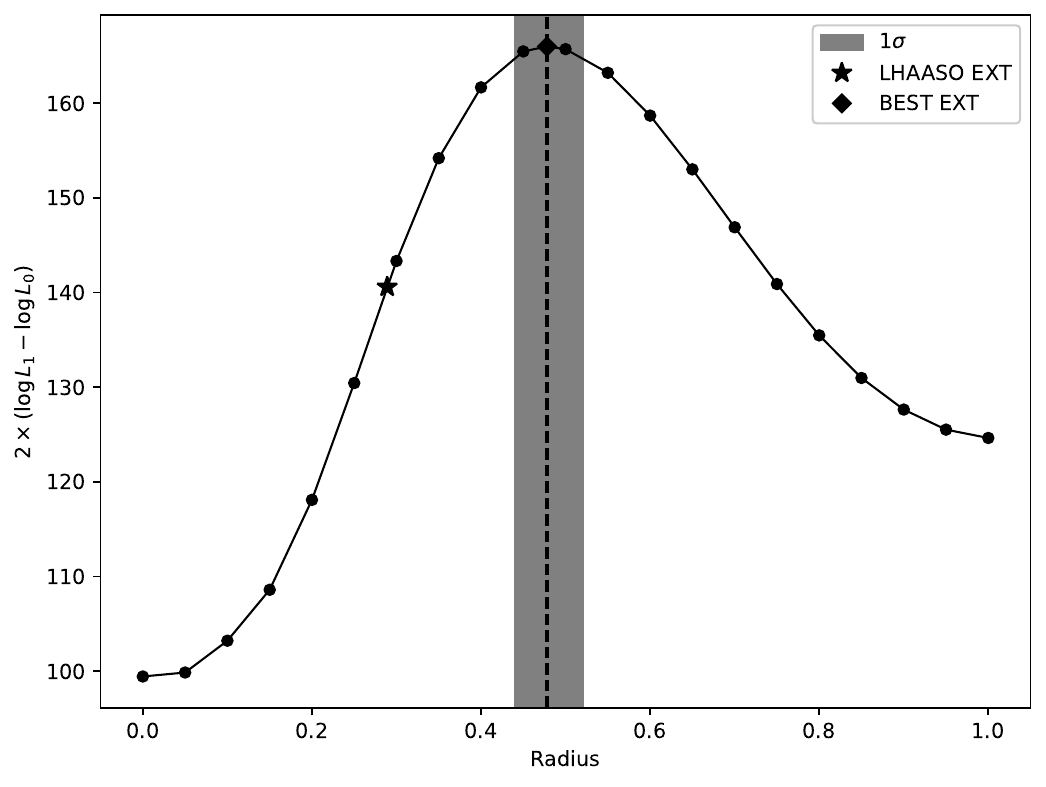}
   \caption{The profile of the TS of the fit improvement when a gaussian disk with gaussian radius of 0.00$^{\circ}$--1.00$^{\circ}$ was considered for the \grslhaaso\ counterpart.}
   \label{fig:ts}
\end{figure}

\begin{figure*}
  \centering
   \includegraphics[width=0.42\textwidth]{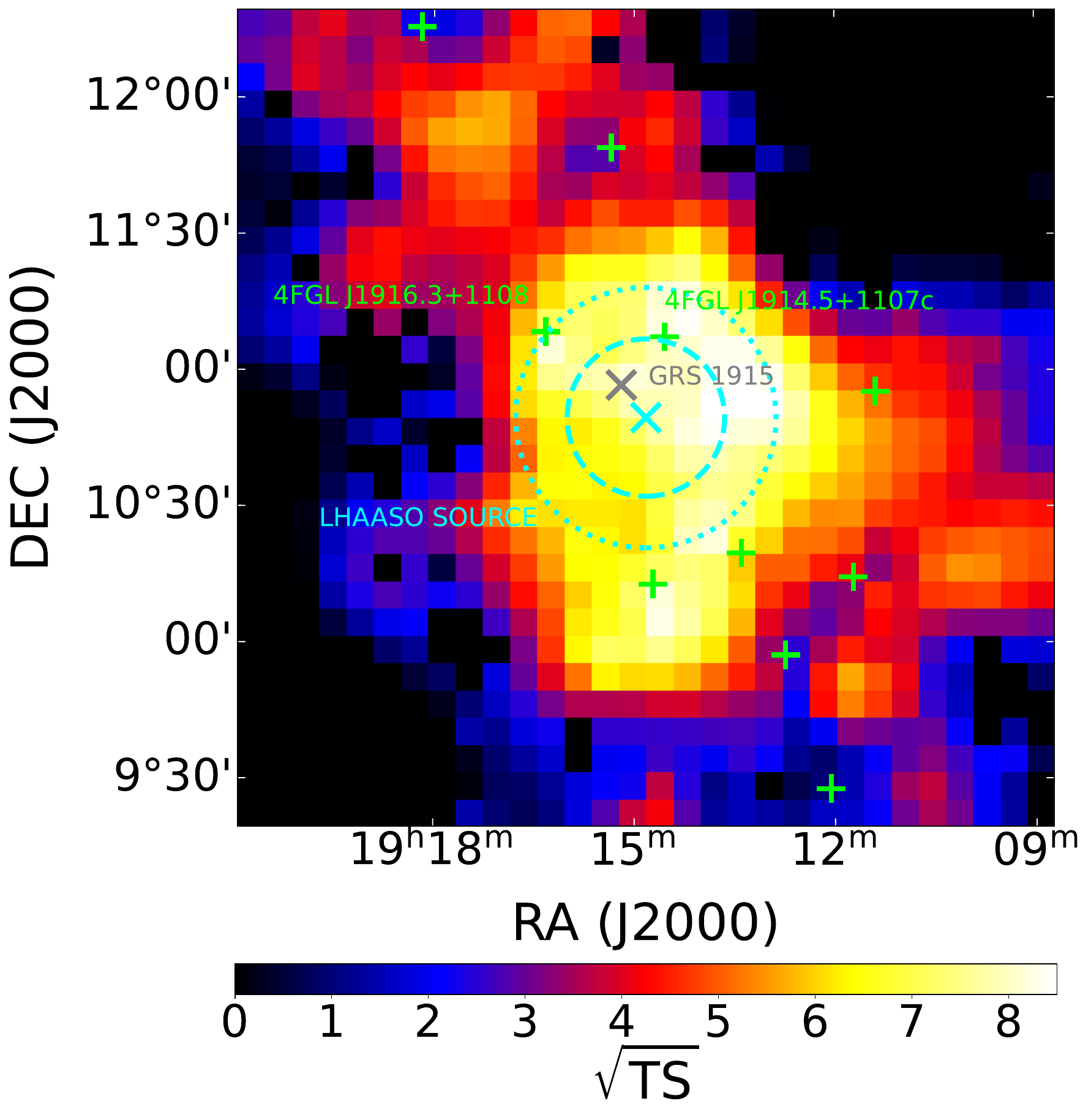}
    \includegraphics[width=0.42\textwidth]{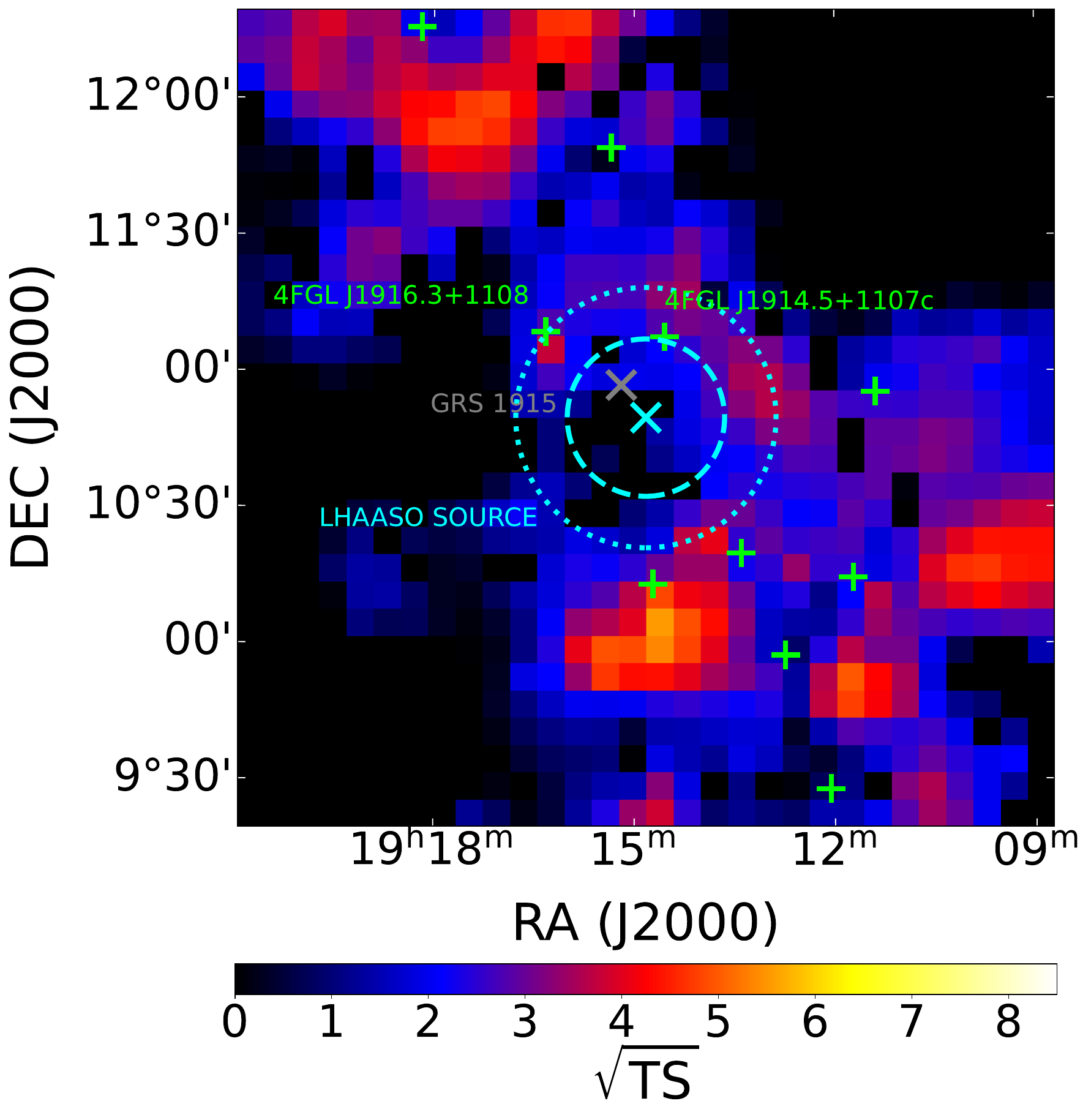}
   \caption{1--500~GeV significance maps for the region of \grslhaaso. All sources in the best source model (4FGL J1916.3 (EXT) $+$  \grslhaaso\ (EXT $0.478^{\circ}$)), except \grslhaaso\ (EXT $0.478^{\circ}$), were removed from the {\it left} panel of the map. \grslhaaso\ (EXT 0.478$^{\circ}$) was further removed in the {\it right} panel. The green pluses mark the positions of the \fermi\ catalog sources, the gray cross marks the position of \grs, the cyan cross marks the center position of \grslhaaso, the cyan dashed and dotted circles mark the extension of \grslhaaso\ derived from LHAASO observation ($0.267^{\circ}$) and our spatial distribution analysis ($0.478^{\circ}$), respectively.
}
   \label{fig:tsmap_check}
\end{figure*}

We removed the \grslhaaso\ counterpart from the source models and instead included a gaussian disk with gaussian radius of $0.00^{\circ}$--$1.00^{\circ}$ (with a initial step of $0.05^{\circ}$ and a finer step of $0.001^{\circ}$ around the best-fit value, where 0.00$^{\circ}$ represents a point source) to describe the \gr\ emission. The center of the gaussian disk was fixed at the center of \grslhaaso.  We performed the likelihood analysis to the 0.1--500~GeV LAT data to assess the source extension by comparing the likelihood values. The TS for extension is evaluated by $-2\ln(\mathcal{L}_\textrm{PS}/\mathcal{L}_\textrm{EXE})$). For consistency, however, we reported the results relative to the baseline source model.
From this analysis, we found that regardless of whether the nearby source 4FGL J1916.3+1108 was modeled as point-like or extended, the maximum TS was obtained when a gaussian disk with a radius of $\sim0.5^{\circ}$ was adopted for \grslhaaso. The TS profile with 4FGL J1916.3+1108 modeled as an extended source is plotted in Fig.~\ref{fig:ts}. The best-fit gaussian radius for the \grslhaaso\ counterpart is $\left(0.478^{+0.043}_{-0.038}\right)^{\circ}$. 

We tested different choices of the center position by adopting the positions of the two new point sources defined in Section~\ref{sec:contamination}, that is the position associated with the micro-quasar GRS 1915+105 and that of PS J1913.4+1050. We found that the maximum TS was consistently obtained at a radius of $0.5^{\circ}$, and the highest TS value was obtained when adopting the center position of GRS 1915–105-LHAASO. We also tested the spatial extension of the two new sources, with GRS 1915–105-LHAASO considered as a extended source with gaussian radius of $0.267^{\circ}$. No significant evidence for extension was found for both of them.

We adopted the best-fit extension of $0.478^{\circ}$ for \grslhaaso\ and updated the analysis results. A spectral index $\Gamma = 2.03 \pm 0.04$ and a 0.1--500~GeV photon flux of $4.4\pm 0.6 \times 10^{-8}\rm\, photons\, cm^{-2}\, s^{-1}$ were obtained for \grslhaaso\ with a TS value of 261. 
The \gr\ spectrum is plotted in the right panel of Fig.~\ref{fig:spectra}, with the flux and TS values provided in Table~\ref{tab:spectra}. The significance map is shown in the left panel of Fig.~\ref{fig:tsmap_check} with only \grslhaaso\ revealed in it, which is further removed in the right panel of Fig.~\ref{fig:tsmap_check}. Compared with the residual significance map in the left panel of Fig.~\ref{fig:tsmap}, the residuals obtained this time were significantly reduced, indicating that the source model including \grslhaaso\ indeed provided a better fit. This is consistent with that inferred from the $\ln\mathcal{L}$ values. 

\subsection{Possible emission contribution from other LHAASO-detected sources}
\label{sec:contribution}

Besides \grslhaaso, three LHAASO-detected sources are located in this region (see Table~\ref{tab:roi_sources}). Although their angular separations from \grslhaaso\ are comparable to or larger than the \fermi-LAT 68\% containment radius of $\sim$0.8$^{\circ}$ at 1~GeV (beyond which significant detections were obtained; see Table~\ref{tab:spectra}), we included them one by one in the source model to assess their potential impact on the results. We found that the best-fit extension is robust and shows no significant change ($<$1$\sigma$), and the \gr\ spectra derived using different source models with or without these three sources are also consistent within uncertainties.
Among these three LHAASO-detected sources, the nearest LHAASO J1912+1010 is an extended source and is spatially coincident with the TeV source HESS J1912+101 reported by the H.E.S.S. Collaboration \citep{aha+08}, and extended emission in the GeV band in this region has been reported (e.g., \cite{zhang+20,zeng+21,li+23}). We also tested the source model considering a Gaussian disk described in the recent paper \citep{li+23} and found that it did not provide a significantly better fit. Nevertheless, the best-fit extension is still robust and shows no significant change ($<$2$\sigma$).
The detailed GeV emission origin around the HESS J1912+101 region is outside the scope of this study, and thus we retained our best-fit source model in the analysis.

\section{Leptonic model}\label{app:leptonic}
High-energy electrons effectively radiate by interacting with different target fields. Most important processes (see in Fig.~\ref{fig:timescale}) include synchrotron radiation caused by interaction with magnetic fields, inverse Compoton (IC) scattering of soft photons, and bremsstrahlung (or free-free collisions with the ambient gas). Unless the density of the ambient gas is very high, \(n_{\mathrm{gas}}\gg100\,\mathrm{cm^{-3}}\), IC process dominates in the TeV band. In extended Galactic sources, the dominant contribution to IC emission comes from the background photon fields, such as CMBR. Therefore, the target field is nearly homogeneous in the source, and the morphologically brightest regions are expected to be associated with the acceleration sites, which in the context of microquasars can be link to the binary system or the jet termination regions. Since LHAASO establishes a shift of the emission centroid to these sites, the leptonic scenarios seem to be disfavored for this source. Nevertheless, below we present a detailed spectral study of the emission in the framework of leptonic scenario.

We approximate the distributions of relativistic particles by the standard models. A broken power law with an exponential cutoff model (BPLEC) is defined as $\dd N/\dd E = A E^{-s_1} [1 + E/E_\textrm{b}]^{(s_1 - s_2)} {\rm exp}(-E/E_{\rm max})$, where $A$ is the normalization constant, $s_1$ and $s_2$ are the power-law indexes, $E_\textrm{b}$ is the break energy, and $E_{\rm max}$ is the cutoff energy. 
We fixed $s_2 = s_1 + 1$ as expected at the transition from slow cooling (or bremsstrahlung dominated cooling) to fast cooling regime under the dominant synchrotron or Thomson losses
The break energy is typically associated with a cooling break and the cutoff energy with the energy at which the acceleration process reaches its limit. 

The main targets for relativistic electrons are the low-temperature photon fields, among which \(2.7\)~K CMB and various infrared fields are most relevant for UHE electrons. Since the parameters of the infrared fields, namely their temperatures and energy densities, are not firmly constrained we consider them as free parameters. For the sake of simplicity, in our modeling we include just a single infrared contribution. Once our fitting yields the preferred properties of the infrared field, we compared it to expected photon field \citep{2017MNRAS.470.2539P}).  Since we extend the analysis to the GeV energy band, we also include bremsstrahlung emission to our modeling. This component doesn't provide any important contribution to VHE and UHE bands, unless the target density is very high, \(n_{\rm gas}\gg100\,\mathrm{cm^{-3}}\),  however may dominate in the GeV band even in sources that have small density of ambient gas. 

\begin{figure}
    \centering
    \includegraphics[width=0.7\linewidth]{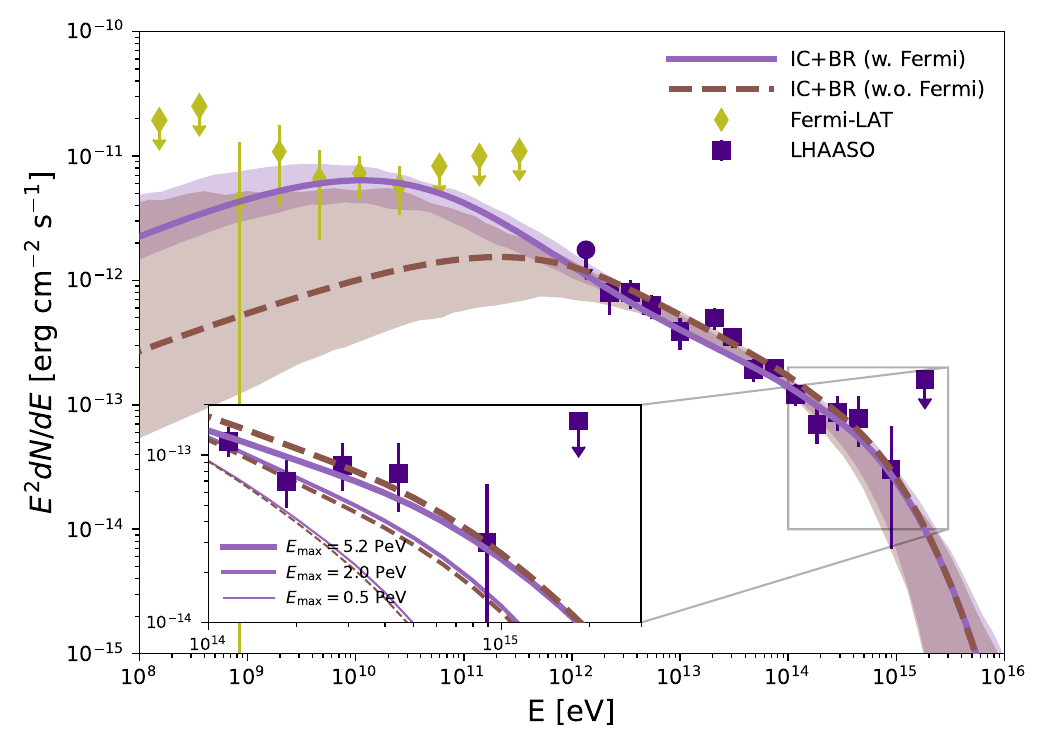}
    \caption{Spectral energy distributions of \grs under leptonic (IC and bremsstrahlung). The dashed curves show the results obtained using LHAASO data only. In the inset figures, we show the leptonic scenario for other two different maximum electron energies, $E_{e,\rm max} = 2\rm~PeV$ and $0.5\rm~PeV$, while keeping all other fitting parameters fixed.}
    \label{fig:GRS1915-SED-fit-leptonic}
\end{figure}

\begin{figure}
    \centering
    \includegraphics[width=0.7\linewidth]{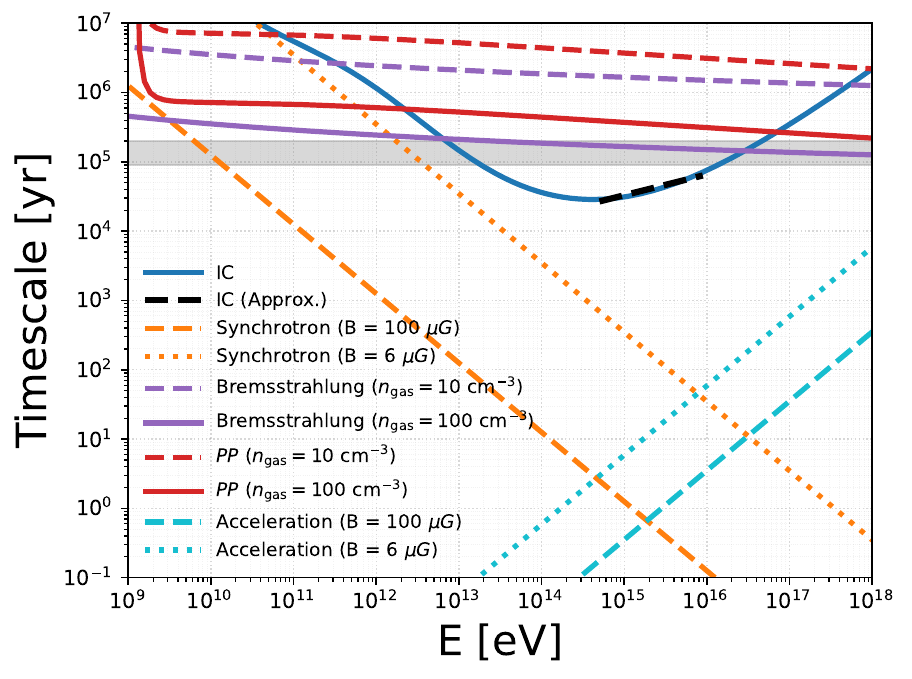}
    \caption{Relevant energy-loss and acceleration timescales for high-energy electrons and protons. The gray shaded region represents the source age range of $[0.09, 0.2]$~Myr. For the acceleration timescale, we adopt $\eta_{\rm acc} = 0.1$. Inverse Compton (IC) cooling includes contributions from the CMB and infrared photons with $T_{\rm IR} \simeq 3050\rm~K$ and energy density $u_{\rm IR} \simeq 1.1\rm~eV~cm^{-3}$.}
    \label{fig:timescale}
\end{figure}

We perform SED fitting for leptonic (i.e., IC + bremsstrahlung) scenario, the results are presented in Fig.~\ref{fig:GRS1915-SED-fit-leptonic}.
The leptonic scenario requires a total electron energy of $W_e \approx 1.9_{-0.6}^{+0.6} \times 10^{48}\rm~erg$ with a spectral index below the break of $s_1 \approx 1.7$. For low-energy electrons, escape and cooling are typically negligible; thus, this index is determined by the acceleration mechanism, and the revealed value is noticeably harder than the canonical value of \(\sim2\).   
The spectral softening of the UHE \gr spectrum is consistent with the Klein-Nishina cutoff, thus the fitting procedure yields a higher value for the cutoff energy, $E_{e, \rm max} \approx 5.2_{-4.3}^{+32.2}\rm~PeV$.
Note that the maximum electron energy is not well constrained; however, the fit indicates a lower limit of approximately $E_{e,\rm max} \gtrsim 1~\mathrm{PeV}$.
To illustrate the role of the cutoff energy, which is a proxy for the efficiency of the acceleration process, we also show calculations for different values of the cutoff energy, $E_{e,\rm max} = 2\rm~PeV$ and $E_{e,\rm max} = 0.5\rm~PeV$ (thin lines in the inset figure, with all other parameters kept fixed).
The parameters fitted for the background infrared photon field, obtained from the leptonic model, correspond to a temperature of $T_{\rm IR} \approx (50 - 3\times 10^4)\,\mathrm{K}$ and an energy density of $u_{\rm IR} \approx (0.24 - 14.3)\,\mathrm{eV\,cm^{-3}}$. These parameters are poorly constrained but their median values are broadly consistent with expected values for infrared background photon fields in the vicinity of \grs\ \citep{2017MNRAS.470.2539P}. 

The spectral break is located at $E_{e, \rm b} \approx 79_{-57}^{+281}\rm~GeV$, which corresponds to the cooling time of $6$~Myr (for the revealed energy density of the photon field and the magnetic field of \(6\,\upmu\mathrm{G}\)). We note that this timescale significantly exceeds the expected age of the source \citep{Motta:2025lgb}, thus the revealed break can hardly be attributed with the cooling break. For high gas density, \(n_{\mathrm{gas}}\sim100\,\mathrm{cm^{-3}}\), the bremsstrahlung cooling might be comparable to the age of the system (see in Fig.~\ref{fig:timescale}), requiring an accurate time-dependent modeling of the particle distribution, which is beyond the scope of this manuscript.

When only LHAASO data are included in the fit, the inferred parameters become $W_e \approx 4.1^{+12.5}_{-3.0} \times 10^{47}\,\mathrm{erg}$, $s_1 \approx 1.8$, $E_{e,\max} \approx 5.1^{+32.3}_{-4.3}\,\mathrm{PeV}$, and $E_{e,b} \approx 1.9^{+5.6}_{-1.5}\,\mathrm{TeV}$. We further note that for a magnetic field of $B = 6~\mu\mathrm{G}$, the synchrotron cooling time of electrons at this energy is of order $\sim 100$ kyr, which is comparable to the characteristic age of the system.

\section{Fitting results from modeling}

\begin{figure*}
    \centering
     \includegraphics[width=0.7\textwidth]{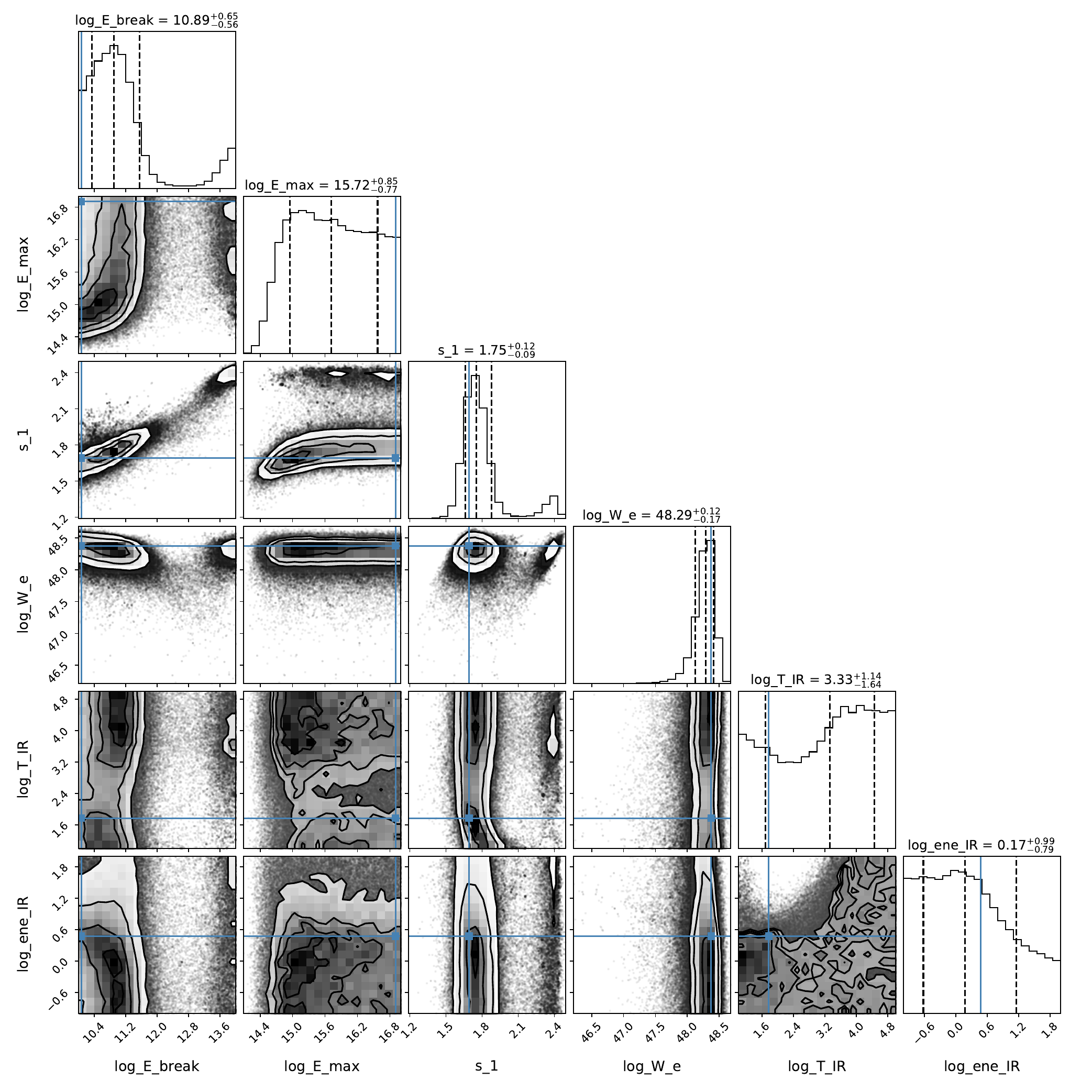}
         \includegraphics[width=0.5 \textwidth]{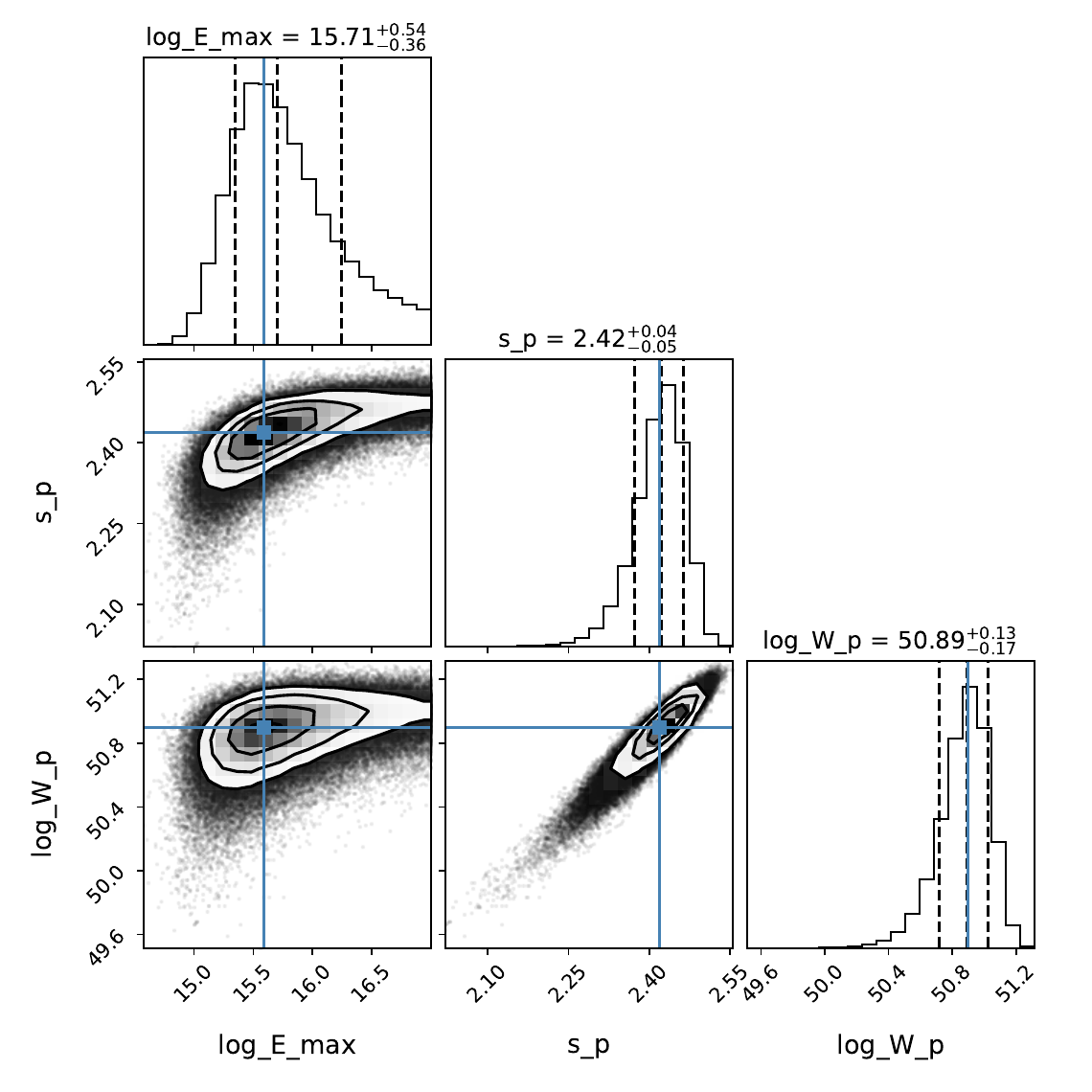}
    \caption{Contour plots of the \gr\ spectrum for the IC (upper panel) and $pp$ (lower panel) models, based on \fermi\ and LHAASO data, respectively.}
    \label{fig:corner-plot-IC-pp}
\end{figure*}

\begin{figure*}
    \centering
     \includegraphics[width=0.7\textwidth]{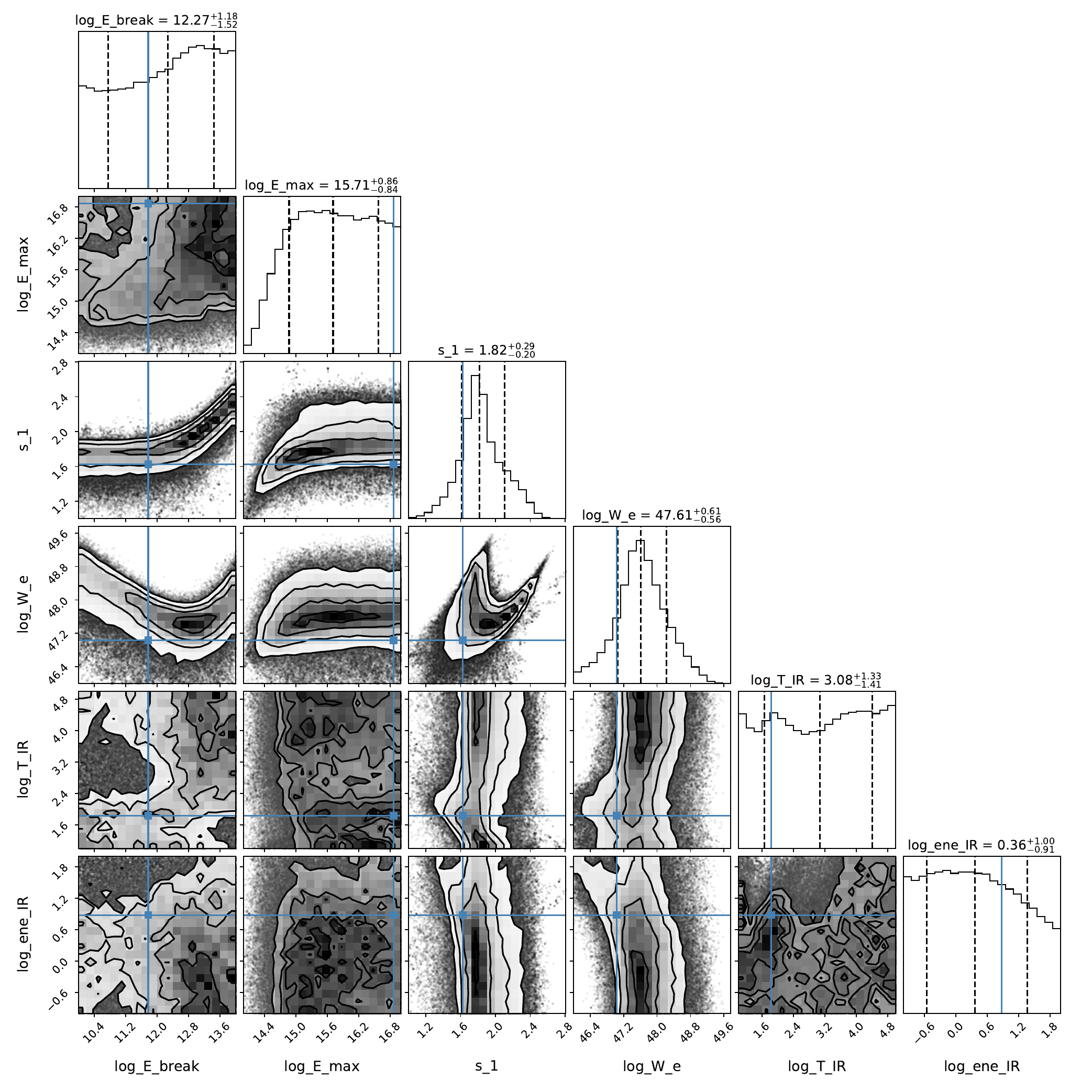}
         \includegraphics[width=0.5 \textwidth]{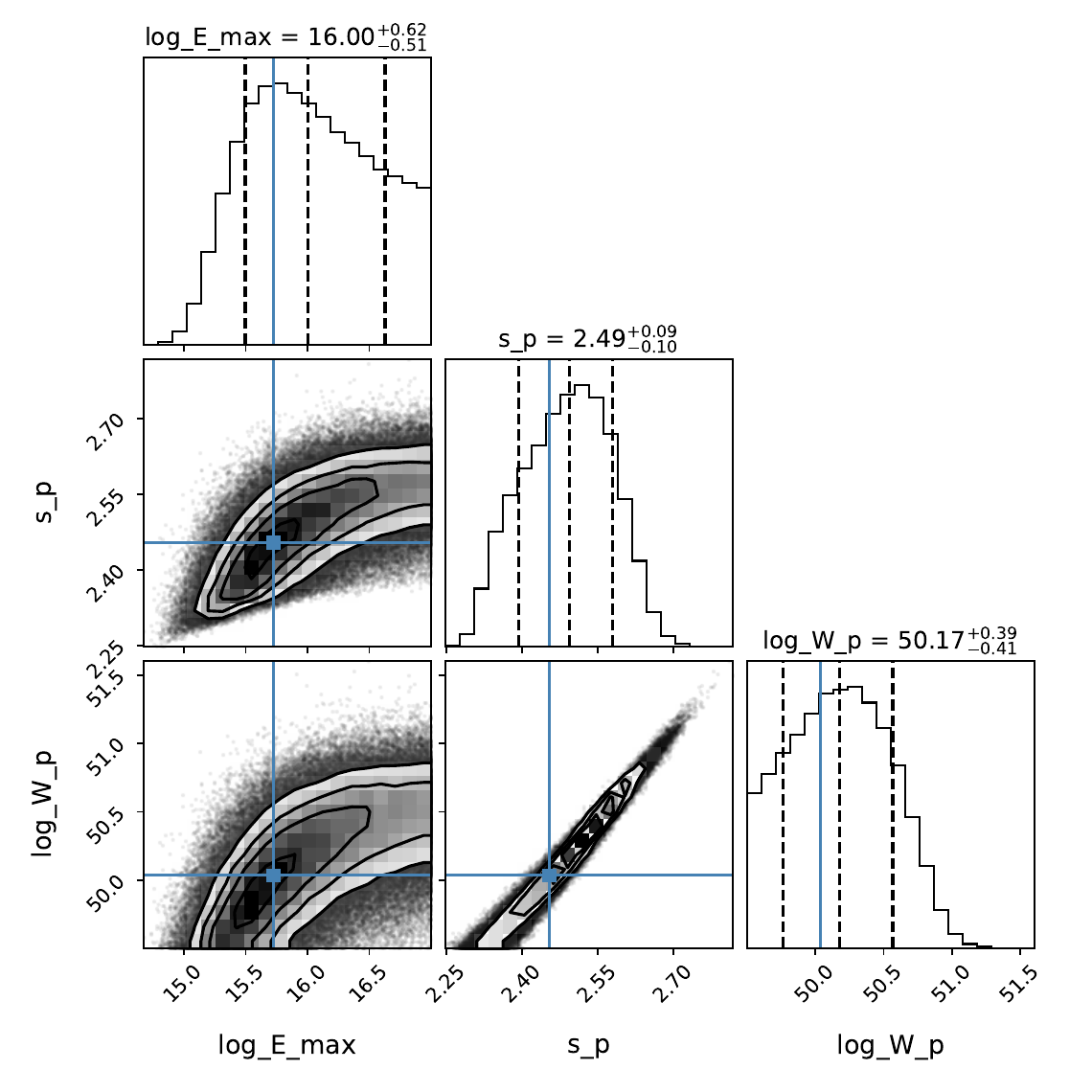}
    \caption{Contour plots of the \gr\ spectrum for the IC (upper panel) and $pp$ (lower panel) models LHAASO data only, respectively.}
    \label{fig:corner-plot-IC-pp-LHAASO}
\end{figure*}

The median parameter values and the adopted prior ranges for the IC and $pp$ models are summarized in Tables~\ref{tab:fit_params_fermi}--\ref{tab:fit_params}, while the corresponding contour plots of the spectral fitting results are shown in Figs.~\ref{fig:corner-plot-IC-pp}--\ref{fig:corner-plot-IC-pp-LHAASO}.
We perform the spectral fitting using {\sc AMES}~\citep{Zhang:2023ewt}, adopting the $pp$ interaction cross section from Ref.~\cite{Kafexhiu:2014cua}. The posterior distributions of the model parameters are sampled using the Markov chain Monte Carlo (MCMC) package {\sc Emcee}~\citep{Foreman-Mackey:2012any}.

\begin{table}[htbp]
\caption{Posterior constraints on the \gr\ spectral parameters derived from joint \fermi\ and LHAASO data, together with the adopted prior ranges. We report the median values with 16th--84th percentile credible intervals.}
\label{tab:fit_params_fermi}
\begin{center}
\renewcommand{\arraystretch}{1.3}
\begin{tabular}{lcc}
\toprule
Symbol & Prior Range & Median $\pm 1\,\sigma$ \\
\midrule
\multicolumn{3}{c}{\textit{IC model}} \\
\midrule
$\log_{10}(E_{\rm b}/\mathrm{eV})$ & $[10.0, 14.0]$ & $10.99_{-0.62}^{+2.58}$ \\
$\log_{10}(E_{\rm max}/\mathrm{eV})$ & $[14.0, 17.0]$ & $15.81_{-0.81}^{+0.83}$ \\
$s_1$ & $[1.0, 4.0]$ & $1.77_{-0.11}^{+0.56}$ \\
$\log_{10}(W_e/\mathrm{erg})$ & $[46.0, 52.0]$ & $48.29_{-0.17}^{+0.12}$ \\
$\log_{10}(T_{\rm IR}/\mathrm{K})$ & $[1.0, 5.0]$ & $3.33_{-1.56}^{+1.14}$ \\
$\log_{10}(u_{\rm IR}/{\rm eV~cm^{-3}})$ & $[-1.0, 2.0]$ & $0.17_{-0.79}^{+0.99}$ \\
\midrule
\multicolumn{3}{c}{\textit{$pp$ model}} \\
\midrule
$\log_{10}(E_{\rm max,p}/\mathrm{eV})$ & $[14.0, 17.0]$ & $15.71_{-0.35}^{+0.54}$ \\
$s_p$ & $[2.0, 3.0]$ & $2.42_{-0.05}^{+0.04}$ \\
$\log_{10}(W_p/\mathrm{erg})$ & $[49.0, 52.0]$ & $49.90_{-0.16}^{+0.13}$ \\
\bottomrule
\end{tabular}
\end{center}
\end{table}

\begin{table}[htbp]
\caption{Best-fit parameters for the \gr\ spectrum derived from LHAASO data only, along with the adopted prior ranges. We report the median values with $1\,\sigma$ uncertainties (16th--84th percentiles).}
\label{tab:fit_params}
\begin{center}
\renewcommand{\arraystretch}{1.3}
\begin{tabular}{lcc}
\toprule
Symbol & Prior Range & Median $\pm 1\,\sigma$ \\
\midrule
\multicolumn{3}{c}{\textit{IC model}} \\
\midrule
$\log_{10}(E_{\rm b}/\text{eV})$ & $[10.0, 14.0]$ & $12.27_{-1.52}^{+1.18}$ \\
$\log_{10}(E_{\rm max}/\text{eV})$ & $[14.0, 17.0]$ & $15.71_{-0.84}^{+0.86}$ \\
$s_1$ & $[1.0, 4.0]$ & $1.82_{-0.20}^{+0.29}$ \\
$\log_{10}(W_e/\text{erg})$ & $[46.0, 52.0]$ & $47.61_{-0.56}^{+0.61}$ \\
$\log_{10}(T_{\rm IR}/\text{K})$ & $[1.0, 5.0]$ & $3.08_{-1.41}^{+1.33}$ \\
$\log_{10}(u_{\rm IR}/{\rm eV~cm^{-3}})$ & $[-1.0, 2.0]$ & $0.36_{-0.91}^{+1.00}$ \\
\midrule
\multicolumn{3}{c}{\textit{$pp$ model}} \\
\midrule
$\log_{10}(E_{\rm max,p}/\mathrm{eV})$ & $[14.0, 17.0]$ & $16.00_{-0.51}^{+0.62}$ \\
$s_p$ & $[2.0, 3.0]$ & $2.49_{-0.10}^{+0.09}$ \\
$\log_{10}(W_p/\mathrm{erg})$ & $[49.0, 52.0]$ & $50.17_{-0.41}^{+0.39}$ \\
\bottomrule
\end{tabular}
\end{center}
\end{table}

\begin{figure*}
    \centering
         \includegraphics[width=0.4 \textwidth]{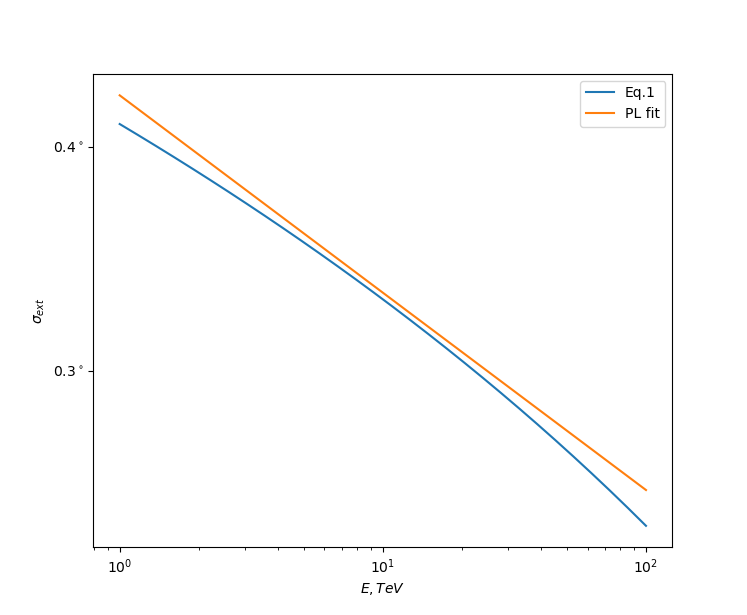}
         \includegraphics[width=0.4405 \textwidth]{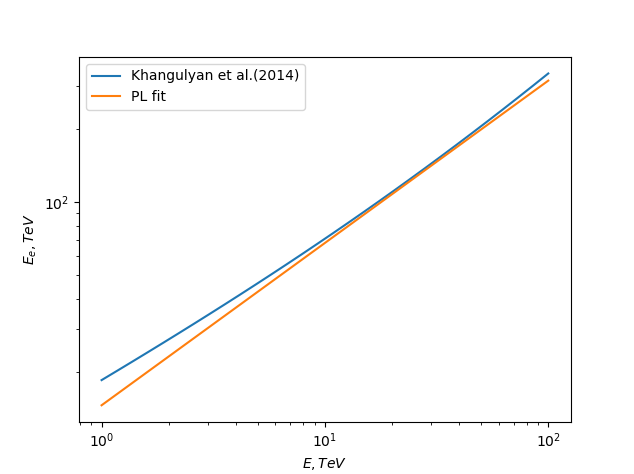}
    \caption{
    Left panel: power-law approximation for the energy-dependent source extension. Right panel: mean energy of parent electrons as a function of energy of the up-scattered photon.}
    \label{fig:simple_fits}
\end{figure*}

\section{Phenomenological implications of the energy-dependent morphology}\label{app:size}
Analysis of the LHAASO data revealed an energy-dependent morphology given by Eq.~\ref{eq:lhaaso_ext}. While the uncertainties of the coefficients in this relation imply a relatively low significance of this discovery, if it is confirmed, it would provide us with important tests for radiative scenarios for the source. The source morphology of gamma-ray sources is determined by the balance of the cooling and transport processes. While higher energy electrons lose energy faster, \(t_{\mathrm{cool}}\propto E_e^{-1}\), the cooling of relativistic protons is approximately energy independent above the threshold energy, \(t_{\mathrm{cool}}\propto \mathrm{const}\). Because of this difference, the expected morphology of hadronic and leptonic sources can be discriminated.  The particle transport might be either advection (which is energy-independent) or diffusion. Since the diffusion coefficient energy dependence, \(D\propto E^\delta\), is not expected to be faster than the one of the electron energy losses, \(\delta<1\), higher energy electrons are expected to be localized closer to the acceleration sites. In contrast, energy-independent cooling of relativistic protons may imply a larger extension of hadronic sources at higher energies. 

Let us first check the possible implication of the energy-dependent morphology for the leptonic scenario. The angular extension of the source revealed with LHAASO, Eq.~\ref{eq:lhaaso_ext}, can be approximated by a power-law function, \(\sigma_{\rm ext} \approx 0.3^\circ\qty(E/25\,\mathrm{TeV})^{-0.11}\). In Fig.~\ref{fig:simple_fits} we show a comparison of the simple analytic approximations for the source energy-dependent extension.  In the energy range relevant for the production of the emission detected with LHAASO, we can approximate the relation between the parent electron energy and the mean photon energy upscattered from CMB as \(E_e\approx 5 E \qty(E/25\,\mathrm{TeV})^{-1/3}\), the relation between electron and \gr energy (for details, see \cite{2014ApJ...783..100K}) is shown in the right panel in Fig.~\ref{fig:simple_fits}. Thus, the energy-dependent extension of electrons should then follow \(\sigma_{\rm ext}\approx 0.3^\circ\qty(E_e/125\,\mathrm{TeV})^{-0.17}\). For the adopted source age, the transport of electrons with \(E_e\gtrsim10\,\mathrm{TeV}\) is dominated by radiative cooling, \(t_{\mathrm{cool}}\approx2.8\times10^{3}\qty(E_e/125\,\mathrm{TeV})^{-1}\qty(B/6\,\mathrm{\upmu G})^{-2}\,\mathrm{yr}\).
We therefore can estimate the phenomenological parameters in the diffusion as:
\begin{equation}
    \chi \qty(\frac{E_e}{1\,\mathrm{PeV}})^{\delta}\approx 250\qty(\frac{E_e}{1\,\mathrm{PeV}})^{2/3}\qty(\frac{B}{6\,\mathrm{\upmu G}})^{3}\,.
\end{equation}
%
Note, that the above power-law index should be considered a rough estimate with uncertainties at the level of 50\% (see the errors in Eq.~\ref{eq:lhaaso_ext}). Provided that, the revealed diffusion might be compatible with a broad range of theoretical models: Bohm, Kolmogorov, or Kraichnan for \(\delta=1\), \(1/3\), and \(1/2\), respectively. Furthermore, for such a fast diffusion, \(\chi\gtrsim10^2\), effects related to the balistic propagation might be important for PeV electrons. 

The energy-dependent morphology expected in the framework of the hadronic scenario is opposite to the one seen with LHAASO. Indeed, higher energy protons can travel larger distances, forming a more extended source. However, if the ballistic effects are important, then the apparent size of the source decreases with energy. Since ballistic effects are important when the diffusion is fast, the confinement time is shorter and the energy requirements for the hadronic scenario increase further. Alternatively, the revealed energy-dependent morphology could also be caused by a recent powerful outburst in the source.


\end{document}

%% file: ApJ20260713.tex
\author{Zhen Cao}
\affiliation{State Key Laboratory of Particle Astrophysics \& Experimental Physics Division \& Computing Center, Institute of High Energy Physics, Chinese Academy of Sciences, 100049 Beijing, China}
\affiliation{University of Chinese Academy of Sciences, 100049 Beijing, China}
\affiliation{Tianfu Cosmic Ray Research Center, 610000 Chengdu, Sichuan,  China}
 
\author{F. Aharonian}
\affiliation{Tianfu Cosmic Ray Research Center, 610000 Chengdu, Sichuan,  China}
\affiliation{University of Science and Technology of China, 230026 Hefei, Anhui, China}
\affiliation{Yerevan State University, 1 Alek Manukyan Street, Yerevan 0025, Armenia }
\affiliation{Max-Planck-Institut for Nuclear Physics, P.O. Box 103980, 69029  Heidelberg, Germany}
 
\author{Y.X. Bai}
\affiliation{State Key Laboratory of Particle Astrophysics \& Experimental Physics Division \& Computing Center, Institute of High Energy Physics, Chinese Academy of Sciences, 100049 Beijing, China}
\affiliation{Tianfu Cosmic Ray Research Center, 610000 Chengdu, Sichuan,  China}
 
\author{Y.W. Bao}
\affiliation{Tsung-Dao Lee Institute \& School of Physics and Astronomy, Shanghai Jiao Tong University, 200240 Shanghai, China}
 
\author{D. Bastieri}
\affiliation{Center for Astrophysics, Guangzhou University, 510006 Guangzhou, Guangdong, China}
 
\author{X.J. Bi}
\affiliation{State Key Laboratory of Particle Astrophysics \& Experimental Physics Division \& Computing Center, Institute of High Energy Physics, Chinese Academy of Sciences, 100049 Beijing, China}
\affiliation{University of Chinese Academy of Sciences, 100049 Beijing, China}
\affiliation{Tianfu Cosmic Ray Research Center, 610000 Chengdu, Sichuan,  China}
 
\author{Y.J. Bi}
\affiliation{State Key Laboratory of Particle Astrophysics \& Experimental Physics Division \& Computing Center, Institute of High Energy Physics, Chinese Academy of Sciences, 100049 Beijing, China}
\affiliation{Tianfu Cosmic Ray Research Center, 610000 Chengdu, Sichuan,  China}
 
\author{W. Bian}
\affiliation{Tsung-Dao Lee Institute \& School of Physics and Astronomy, Shanghai Jiao Tong University, 200240 Shanghai, China}
 
\author{J. Blunier}
\affiliation{APC, Universit\'e Paris Cit\'e, CNRS/IN2P3, CEA/IRFU, Observatoire de Paris, 119 75205 Paris, France}
 
\author{A.V. Bukevich}
\affiliation{Institute for Nuclear Research of Russian Academy of Sciences, 117312 Moscow, Russia}
 
% ...

\collaboration{330}{The LHAASO Collaboration}